\documentclass[]{aastex631}

\usepackage{lipsum}
\usepackage{tabularx}
\usepackage{array}
\usepackage{layout}
\usepackage{booktabs}
\usepackage{xspace}
\usepackage{upgreek}

\usepackage[T1]{fontenc}
\usepackage{textcomp}
\usepackage{lmodern}

\newcommand{\ronefourseven}{FRB~20240114A\xspace}

\newcommand{\evnpointA}{$\alpha = 21^{\mathrm{h}}27^{\mathrm{m}}39\fs888$, $\delta = +04\degr21\arcmin00\farcs36$ (J2000)}

\newcommand{\evnpointB}{$\alpha = 21^{\mathrm{h}}27^{\mathrm{m}}39\fs8367$, $\delta = +04\degr19\arcmin46\farcs2333$ (J2000)}

\newcommand{\evndelaypos}{$\alpha = 21^{\mathrm{h}}27^{\mathrm{m}}39\fs9$, $\delta = +04\degr19\arcmin43\farcs4$ (J2000)}

\newcommand{\evnatelpos}{$\alpha = 21^{\mathrm{h}}27^{\mathrm{m}}39\fs835$, $\delta = +04\degr19\arcmin45\farcs634$}

\newcommand{\evnbestpos}{$\alpha = 21^{\mathrm{h}}27^{\mathrm{m}}39\fs835$, $\delta = +04\degr19\arcmin45\farcs668$ (J2000, ICRF)}

\begin{document}

\title{A Hyperactive Fast Radio Burst Pinpointed in an SMC-like Satellite Host Galaxy}

\shorttitle{A Repeating FRB in a Small Magellanic Cloud Analog}

\correspondingauthor{Mohit Bhardwaj \& Mark Snelders}
\email{work.mohitbhardwaj@gmail.com, snelders@astron.nl}
\equalcontributionsymbol{Co-first authors of this manuscript.}

\author[0000-0002-3615-3514]{M.~Bhardwaj\textsuperscript{\textdagger}}
\affiliation{McWilliams Center for Cosmology \& Astrophysics, Department of Physics, Carnegie Mellon University, Pittsburgh, PA 15213, USA}

\author[0000-0001-6170-2282]{M.~P.~Snelders\textsuperscript{\textdagger}}
\affiliation{ASTRON, Netherlands Institute for Radio Astronomy, Oude Hoogeveensedijk 4, 7991 PD Dwingeloo, The Netherlands}
\affiliation{Anton Pannekoek Institute for Astronomy, University of Amsterdam, Science Park 904, 1098 XH, Amsterdam, The Netherlands}

\author[0000-0003-2317-1446]{J.~W.~T.~Hessels}
\affiliation{Department of Physics, McGill University, 3600 rue University, Montréal, QC H3A 2T8, Canada}
\affiliation{Trottier Space Institute, McGill University, 3550 rue University, Montréal, QC H3A 2A7, Canada}
\affiliation{ASTRON, Netherlands Institute for Radio Astronomy, Oude Hoogeveensedijk 4, 7991 PD Dwingeloo, The Netherlands}
\affiliation{Anton Pannekoek Institute for Astronomy, University of Amsterdam, Science Park 904, 1098 XH, Amsterdam, The Netherlands}

\author[0000-0001-6150-2854]{A.~Gil~de~Paz}
\affiliation{Departamento de Física de la Tierra y Astrofísica, Facultad de CC. Físicas, Universidad Complutense de Madrid, E-28040, Madrid, Spain}
\affiliation{Instituto de Física de Partículas y del Cosmos IPARCOS, Facultad de CC. Físicas, Universidad Complutense de Madrid, E-28040 Madrid, Spain}

\author[0000-0003-3460-506X]{S.~Bhandari}
\affiliation{SKA Observatory (SKAO), Science Operations Centre, CSIRO ARRC, Kensington WA 6151, Australia}

\author[0000-0001-9814-2354]{B.~Marcote}
\affiliation{Joint Institute for VLBI ERIC, Oude Hoogeveensedijk 4, 7991~PD Dwingeloo, The Netherlands}
\affiliation{ASTRON, Netherlands Institute for Radio Astronomy, Oude Hoogeveensedijk 4, 7991 PD Dwingeloo, The Netherlands}

\author[0000-0002-8139-8414]{A.~Kirichenko}
\affiliation{Instituto de Astronomía, Universidad Nacional Autónoma de México, Apdo. Postal 877, Ensenada, Baja California 22800, México}

\author[0000-0001-9381-8466]{O.~S.~Ould-Boukattine}
\affiliation{ASTRON, Netherlands Institute for Radio Astronomy, Oude Hoogeveensedijk 4, 7991 PD Dwingeloo, The Netherlands}
\affiliation{Anton Pannekoek Institute for Astronomy, University of Amsterdam, Science Park 904, 1098 XH, Amsterdam, The Netherlands}

\author[0000-0001-6664-8668]{F.~Kirsten}
\affiliation{Department of Space, Earth and Environment, Chalmers University of Technology, Onsala Space Observatory, 439 92, Onsala, Sweden}
\affiliation{ASTRON, Netherlands Institute for Radio Astronomy, Oude Hoogeveensedijk 4, 7991 PD Dwingeloo, The Netherlands}

\author[0000-0002-1727-1224]{E.~K.~Bempong-Manful}
\affiliation{Jodrell Bank Centre for Astrophysics, Dept.\ of Physics \& Astronomy, University of Manchester, Manchester M13 9PL, UK}
\affiliation{School of Physics, University of Bristol, Tyndall Avenue, Bristol BS8 1TL, UK}

\author[0000-0003-3655-2280]{V.~Bezrukovs}
\affiliation{Engineering Research Institute Ventspils International Radio Astronomy Centre (ERI VIRAC) of Ventspils University of Applied Sciences, Inzenieru street 101, Ventspils, LV-3601, Latvia}

\author[0000-0002-0963-0223]{J.~D.~Bray}
\affiliation{Jodrell Bank Centre for Astrophysics, Dept.\ of Physics \& Astronomy, University of Manchester, Manchester M13 9PL, UK}

\author[0000-0002-3341-466X]{S.~Buttaccio}
\affiliation{INAF-Istituto di Radioastronomia, Via Gobetti 101, 40129, Bologna, Italy}
\affiliation{INAF-Osservatorio Astrofisico di Catania, via Santa Sofia 78, I-95123, Catania, Italy}

\author[0000-0002-5924-3141]{A.~Corongiu}
\affiliation{INAF-Osservatorio Astronomico di Cagliari, via della Scienza 5, I-09047, Selargius (CA), Italy}

\author[0000-0002-9812-2078]{R.~Feiler}
\affiliation{Institute of Astronomy, Faculty of Physics, Astronomy and Informatics, Nicolaus Copernicus University, Grudziadzka 5, PL-87-100 Toru\'n, Poland}

\author[0000-0003-4056-4903]{M.~P.~Gawro\'nski}
\affiliation{Institute of Astronomy, Faculty of Physics, Astronomy and Informatics, Nicolaus Copernicus University, Grudziadzka 5, PL-87-100 Toru\'n, Poland}
\affiliation{Anton Pannekoek Institute for Astronomy, University of Amsterdam, Science Park 904, 1098 XH, Amsterdam, The Netherlands}

\author[0000-0002-8657-8852]{M.~Giroletti}
\affiliation{INAF-Istituto di Radioastronomia, Via Gobetti 101, 40129, Bologna, Italy}

\author[0000-0002-5794-2360]{D.~M.~Hewitt}
\affiliation{Anton Pannekoek Institute for Astronomy, University of Amsterdam, Science Park 904, 1098 XH, Amsterdam, The Netherlands}

\author[0000-0002-3669-0715]{M.~Lindqvist}
\affiliation{Department of Space, Earth and Environment, Chalmers University of Technology, Onsala Space Observatory, 439 92, Onsala, Sweden}

\author[0000-0002-1482-708X]{G.~Maccaferri}
\affiliation{INAF-Istituto di Radioastronomia, Via Gobetti 101, 40129, Bologna, Italy}

\author[0000-0003-1936-9062]{A.~Moroianu}
\affiliation{Anton Pannekoek Institute for Astronomy, University of Amsterdam, Science Park 904, 1098 XH, Amsterdam, The Netherlands}

\author[0000-0003-0510-0740]{K.~Nimmo}
\affiliation{MIT Kavli Institute for Astrophysics and Space Research, Massachusetts Institute of Technology, 77 Massachusetts Ave, Cambridge, MA 02139, USA}

\author[0000-0002-5195-335X]{Z.~Paragi}
\affiliation{Joint Institute for VLBI ERIC, Oude Hoogeveensedijk 4, 7991~PD Dwingeloo, The Netherlands}

\author[0000-0003-2422-6605]{W.~Puchalska}
\affiliation{Institute of Astronomy, Faculty of Physics, Astronomy and Informatics, Nicolaus Copernicus University, Grudziadzka 5, PL-87-100 Toru\'n, Poland}

\author[0000-0002-9786-8548]{N.~Wang}
\affiliation{Xinjiang Astronomical Observatory, CAS, 150 Science 1-Street, Urumqi, Xinjiang 830011, China}

\author[0000-0001-7361-0246]{D.~Williams-Baldwin}
\affiliation{Jodrell Bank Centre for Astrophysics, Dept.\ of Physics \& Astronomy, University of Manchester, Manchester M13 9PL, UK}

\author[0000-0002-5381-6498]{J.P.~Yuan}
\affiliation{Xinjiang Astronomical Observatory, CAS, 150 Science 1-Street, Urumqi, Xinjiang 830011, China}

\begin{abstract}
Precise localizations of fast radio bursts (FRBs) are essential for uncovering their host galaxies and immediate environments. We present the milliarcsecond-precision European VLBI Network localization of \ronefourseven, a hyperactive repeating FRB, achieving $\lesssim$$90\times30$\,mas ($1\sigma$) accuracy. This precision places the burst $0.5$\,kpc from the nucleus of its low-metallicity star-forming dwarf host at a spectroscopic redshift of $z = 0.130287$. Our Gran Telescopio CANARIAS spectroscopic follow-up reveals that the dwarf FRB host is gravitationally bound to a more massive, star-forming spiral galaxy. This establishes the first known instance of an FRB residing in a satellite galaxy within a larger galactic system. This configuration, analogous to the Small Magellanic Cloud orbiting the Milky Way (but at a lower overall mass scale), expands the known diversity of FRB host environments and offers important insights for interpreting seemingly ``hostless'' or highly offset FRBs. Furthermore, our detailed dispersion measure (DM) budget analysis indicates that the dominant contribution to \ronefourseven's DM likely originates from the foreground galaxy halos. This finding addresses the anomalously high DM observed for this FRB and underscores the significant role of intervening foreground structures in shaping observed FRB DMs, which is important for accurate FRB-based cosmological measurements. Our results highlight the importance of deep, high-resolution optical/infrared observations (e.g., with the \textit{Hubble} or \textit{James Webb Space Telescopes}) to fully leverage our precise radio localization and probe the immediate astrophysical birthplaces of FRB progenitors within these complex galactic systems.
\end{abstract}


\section{Introduction} \label{sec:intro}

Fast radio bursts (FRBs) are brief flashes of radio waves
originating from extragalactic distances \citep{Lorimer_2007_Sci}. Given the observed energetics and timescales of FRBs they must come from compact objects \citep[see, e.g.,][for a review]{Petroff_2022_A&ARv}, but their exact origin(s) and emission mechanism(s) remain unclear. A small fraction ($\sim$$2$\,\%) of the FRB population has been observed to repeat \citep[e.g.,][]{Spitler_2016_Natur,Chime/FrbCollaboration_2023_ApJ}, with more than one burst coming from the same position on the sky with the same dispersion measure (DM), which shows that at least a part of the population requires a central engine that can produce bursts over the course of at least months to years. The magnetically powered neutron stars known as `magnetars' are a popular explanation for the FRB phenomenon --- both because of an energetic FRB-like event from a known Galactic magnetar, SGR~1935+2154 \citep{Bochenek_2020_Natur,CHIME/FRBCollaboration_2020_Natur_SGR1935}, and the large number of theoretical models that describe the exact mechanisms by which magnetars can generate impulsive radio emission \citep[see, e.g.,][]{Zhang_2020_Natur}. However, at this point it is unclear if repeating FRBs and apparent non-repeaters share a common origin \citep[e.g.,][]{Pleunis_2021_ApJ}. Likewise, it is debatable whether the population as a whole is due to a single progenitor type and emission mechanism (see e.g., \citet{Kirsten_2022_Natur} and \citet{hewitt_2023_mnras} for discussions about different progenitors and emission mechanisms, respectively).

Valuable insights can be gained by analyzing the burst properties of FRBs themselves, such as; burst durations \citep[e.g.,][]{snelders_2023_natas}, their scintillation bandwidths and scattering timescales \citep[e.g.,][]{nimmo_2025_nature}, their (time variable) Faraday rotation measure \citep[RM; e.g.,][]{Michilli_2018_Natur} and dispersion measure \citep[e.g.,][]{Hessels_2019_ApJL}, energetics \citep[e.g.,][]{ouldboukattine_2024_arxiv} and (de)polarization \citep[e.g.,][]{gopinath_2024_mnras} --- all of these properties reveal information about the possible emission mechanism(s) and propagation effects. Besides studying the properties of the bursts themselves, the origin(s) of FRBs can be investigated by identifying their host galaxies and local environments. Localizations of apparently one-off FRBs can be achieved by detecting them with wide-field-of-view interferometric arrays operating at high time resolution, such as the Deep Synoptic Array \citep[DSA; e.g.,][]{Sharma_2024_Nature}, the Australian Square Kilometre Array Pathfinder \citep[ASKAP; e.g.,][]{Gordon_2024_APJL}, and the Canadian Hydrogen Intensity Mapping Experiment FRB system (CHIME/FRB) and its Outrigger telescopes \citep[e.g.,][]{Shah_2025_ApJL}. To determine the local neighbourhood of an FRB source, within its host, a localization precision on the milliarcsecond (mas) level is needed. This can only be achieved using very long baseline interferometry (VLBI), with baselines of hundreds to thousands of kilometres. The CHIME Outriggers system is now starting to provide $\sim$$100$\,mas FRB positions \citep{andrew_2025_atel,chimefrb_2025_apjl_rbfloat}. The most precise localizations of repeaters, and their sometimes associated compact persistent radio sources (PRS), have been achieved with the European VLBI Network (EVN), as we do in this Letter \citep{Marcote_2017_ApJL,Marcote_2020_Natur,Kirsten_2022_Natur,Nimmo_2022_ApJL,Bhandari_2023_APJL_Dec,Hewitt_2024_MNRAS,hewitt_2024_mnras_dec,moroianu_2025_arxiv}. 

Different degrees of localization precision, provide increasing clarity on an FRB's environment. While arcsecond localization is normally sufficient to identify a robust host galaxy association \citep{eftekhari_2017_apj}, sub-arcsecond localization is required to pinpoint an FRB's local galactic neighbourhood \citep{Mannings_2021_ApJ} --- e.g., associating it with a spiral arm or star-forming region. At the mas-level, an FRB could even be associated with a specific stellar counterpart, precisely situated with respect to nearby star-forming regions \citep{Tendulkar_2021_ApJL} or globular cluster \citep{Kirsten_2022_Natur}, and/or associated with a compact persistent radio source \citep[e.g.,][]{Marcote_2017_ApJL}. Here, and in the literature, an FRB's `local environment', is used to mean the galactic neighborhood ($\sim$kpc scale) or immediate surroundings of the source (few to tens of pc).

Among repeating FRBs, a rare subset of a few sources show orders-of-magnitude higher burst rate than other repeaters. They are colloquially called `hyperactive' repeaters, with two prominent examples being FRB~20201124A \citep[e.g.,][]{Xu_2022_Natur} and FRB~20220912A \citep[e.g.,][]{Konijn_2024_MNRAS}, both discovered by CHIME/FRB. One of these hyperactive repeating FRB sources is another CHIME/FRB discovery, \ronefourseven \citep{Shin_2024_ATel,shin_2025_arxiv}. Since its discovery, CHIME/FRB has detected a handful of bursts from this source\footnote{\url{https://www.chime-frb.ca/repeaters/FRB20240114A}}, while thousands more have been detected by various telescopes worldwide (e.g., \citealt{ouldboukattine_2025_atel}, \citealt{kumar_2024_apj}, \citealt{xie_2025_apjs} and \citealt{Zhang_2024_ATel_March}).

Using MeerKAT, \citet{Tian_2024_MNRAS} were able to robustly localize \ronefourseven, with an uncertainty of $1.4\arcsec$, to a galaxy named J212739.84+041945.8 in the Sloan Digital Sky Survey (SDSS). \citet{Bhardwaj_2024_ATel} used the Gran Telescopio CANARIAS (GTC) to determine that the host is a dwarf star-forming galaxy at a redshift of $z = 0.13$. Co-located continuum radio emission was found using the upgraded Giant Metrewave Radio Telescope \citep[uGMRT;][]{bhusare_arxiv_2024} and an apparent compact PRS was found at $\sim$$5$\,GHz using the Very Long Baseline Array \citep[VLBA;][]{bruni_2025_aa}. Using the GTC, \citet{chen_ApJ_2025} find a redshift that is consistent with \citet{Bhardwaj_2024_ATel} and show that the host of \ronefourseven is located behind a galaxy cluster.

In this Letter, we present a $\sim$$90\times30$\,mas ($1\sigma$) localization of \ronefourseven using the EVN, which is $\sim$$3000$ times more precise than the previous localization of this source in terms of sky area. We complement this precise position with new, deep spectroscopic observations of the host galaxy and surrounding field using the GTC. We show that the host is a low-metallicity dwarf galaxy that is a satellite of a more massive central galaxy --- analogous to the Small Magellanic Cloud and Milky Way system. The radio and optical observations are described in \S\,\ref{subsec:evenobs} and \S\,\ref{subsec:GTCobs}, respectively. The data analysis and results are presented in \S\,\ref{sec:results} and \S\,\ref{sec:intervening_dm}. Finally, we discuss the implications of these observations in \S\,\ref{sec:discussion_conclusion}. Throughout this analysis, we adopt Planck 2015 cosmological parameters \citep[$H_0 = 67.8$\,km\,s$^{-1}$\,Mpc$^{-1}$, $\Omega_m = 0.308$, $\Omega_\Lambda = 0.692$;][]{2016A&A...594A..13P}.

\section{Observations} \label{sec:obs}
\subsection{EVN Observations} \label{subsec:evenobs}
We observed \ronefourseven at two epochs in February 2024 using an {\it ad-hoc} sub-array of EVN dishes (EVN-Lite mode), as part of the PRECISE project (Pinpointing REpeating ChIme Sources with EVN dishes; PI:~F.~Kirsten). Both observing runs were carried out between $1254$--$1510$\,MHz in a mixed frequency setup (see Appendix~\ref{appendix:radio_observations}). All stations recorded raw voltages as $2$-bit samples, storing both left and right circular polarizations in VDIF format \citep{Whitney_2010_ivs}.

Epoch~1 (PRECISE code PR318A; EVN code EK056A) was conducted on 15 February 2024 (MJD 60355) with the following stations: Effelsberg (Ef; Germany), Toru\'{n} (Tr; Poland), Onsala (O8; Sweden), Westerbork RT-1 (Wb; The Netherlands), Noto (Nt; Italy), Irbene (Ir; Latvia) and six e-MERLIN stations (Cambridge (Cm), Darnhall (Da), Defford (De), Knockin (Kn), Pickmere (Pi), and Jodrell Bank Mark II (Jm); United Kingdom). During Epoch~1, the antenna pointing position towards \ronefourseven was \evnpointA, which is the CHIME/FRB baseband position of the source \citep{Shin_2024_ATel}.

Epoch~2 (PR319A/EK056B) was conducted on 20 February 2024 (MJD 60360) with the same stations as in Epoch~1, but with the addition of the Tianma 65-metre Telescope (T6; China), which participated for only part of the observation due to elevation limits (see Appendix~\ref{appendix:radio_observations}). The antenna pointing position for Epoch~2 was \evnpointB, derived from the MeerKAT burst positions reported in \citet{Tian_2024_ATel}. 

We used phase-referencing with a non-conservative cycle time of $\sim$$8$\,minutes, where $\sim$$6$\,minute scans on \ronefourseven are interleaved with $\sim$$2$\,minute phase-calibrator scans on J2125+0441 ($0.65\degr$ away from \ronefourseven; quoted times are including telescope slewing, see Appendix~\ref{appendix:radio_observations}). The $\sim$$8$\,minute cycle time is relatively long and was chosen to maximize the time on \ronefourseven. Additionally, we observed J2120+0246 roughly twice per hour and used it as a check-source to verify the astrometric solutions. The source J1829+4844 was used as a fringe finder and bandpass calibrator and the flux calibrator 3C286 (J1331+3030) was observed with the e-MERLIN stations. Various pulsars (Appendix~\ref{appendix:radio_observations}) were observed to verify the data quality, frequency setup and burst-search pipeline. We achieved a total on-source time with the Effelsberg telescope (the largest dish in the array\footnote{\url{https://www.evlbi.org/sites/default/files/shared/EVNstatus.txt}}) of $200$\,minutes during Epoch~1 and $202$\,minutes during Epoch~2.

\subsection{GTC Observations} \label{subsec:GTCobs}

We obtained the spectra of the host and central galaxies (\S\,\ref{subsec:optical_analysis}) using the Optical System for Imaging and low-intermediate Resolution Integrated Spectroscopy (OSIRIS+) instrument installed at the 10.4m- GTC during three observing runs (programs GTCMULTIPLE2G-24A and GTCMULTIPLE2C-25AMEX).

The initial observations were performed on July 4, 2024, under clear conditions with seeing of $\sim$$0.7\arcsec$. We will refer to these observations as the 2024 run. We acquired $4\times800$\,second exposures using the R1000B grism, which covers the spectral range $3630$--$7000$\,\AA. The two galaxies were exposed in a single long slit with the $1.2\arcsec$ width, providing a spectral resolution of about $10$\,\AA.

Two additional data sets were obtained on June 19 and 22, 2025, under clear conditions, using the higher resolution grism R2500R, with the spectral range $5575$--$7685$\,\AA, to get more precise redshifts of the galaxies (see \S\,\ref{subsec:optical_analysis}). This will be referred to as the 2025 run. In this case, each set contained $4\times1200$\,second exposures, and seeing varied between $0.8\arcsec$ and $0.9\arcsec$ during both observations. The slit orientation in the first set was identical to that in the 2024 observation, covering the two galaxies simultaneously. The second observing night targeted only the central galaxy, with the slit placed along the major axis. We used the $0.8\arcsec$ slit width, providing a spectral resolution of $\sim$$3$\,\AA.

Data reduction was performed with the standard Image Reduction and Analysis Facility (IRAF) routines. The wavelength solution was achieved using a set of HgAr and Ne arc lamps, with a resulting rms of $\lesssim 0.2$\,\AA\ for the R1000B grism (2024 run) and $\lesssim 0.02$\,\AA\ for the R2500R grism (2025 run). In the case of the observations taken on the first night of the 2025 run, we generated two sets of processed spectra of the FRB host (named 2025A \& 2025B) using two different sets of arc-lamp observations that were acquired before and after observing the science target, respectively. The reduced spectrum taken along the major axis of the central galaxy is referred to as 2025C. The flux was calibrated using the Feige~66 and Feige~110 spectrophotometric standards \citep{oke}.

\section{Analysis and Results}\label{sec:results}

\subsection{Radio Analysis and Results}\label{subsec:radio_analysis}

\subsubsection{Search for Bursts}
\label{subsubsec:burstsearch}
The raw voltage data from the Effelsberg telescope were transported to a dedicated server at Onsala Space Observatory (Sweden) to be searched for FRBs. The data were converted to Stokes~I (intensity) filterbank files \citep{Lorimer_2011_ascl}, using \texttt{digifil} \citep{vanStraten_2011_PASA}. The resulting files had a time and frequency resolution of $64$\,$\upmu$s and $62.5$\,kHz, respectively. The intensity data were searched for FRBs with \texttt{Heimdall} \citep{Heimdall}, using a DM range of $527.7 \pm 50$\,pc\,cm$^{-3}$. This DM is centered around the detection DM of CHIME/FRB \citep{Shin_2024_ATel} and fully covers any potential change in DM for this source, which does not exceed $\sim$$1.3$\,pc\,cm$^{-3}$ \citep{shin_2025_arxiv}. FRB-candidates identified as such by \texttt{Heimdall} were further analyzed by \texttt{FETCH} \citep{Agarwal_2020_MNRAS}, a deep-learning-based transient classifier. We manually inspect all candidates for which \texttt{FETCH} assigned a $\geq0.5$ probability that the candidate was of astrophysical origin. The full details of the pipeline are described in \citet{Kirsten_2021_NatAs, Kirsten_2022_Natur}. We detected $6$ bursts in Epoch~1 and $13$ bursts in Epoch~2. These bursts are labeled A$01$--A$07$ and B$01$--B$13$ for the first and second epoch, respectively. Burst A$02$ was initially picked up by the pipeline but during the analysis the burst was deemed too weak to be confidently labeled as a real FRB and was therefore removed from the sample.

\subsubsection{Interferometric Data}
\label{subsubsec:correlation}
After the bursts were detected, the baseband data of all participating stations were transferred to the Joint Institute for VLBI ERIC (JIVE; The Netherlands) for correlation with the FX Software Correlator \texttt{SFXC} \citep{Keimpema_2015_ExA}. The calibrator data were correlated using $2$\,s integrations and $0.5$\,MHz channel widths ($64$\,channels per $32$\,MHz subband). Using the delay-mapping technique described in \cite{Marcote_2017_ApJL,Marcote_2020_Natur}, we derived a rough (arcsecond-level precision) position, \evndelaypos, for the brightest burst of Epoch~1 (A$04$). The data of the brightest burst were re-correlated at that position. Those data were reduced, calibrated and imaged with \texttt{CASA} \citep{CASA} --- following the same calibration strategy as described in \citet{Bhandari_2023_APJL_Dec} --- leading to a best position of that single burst of \evnatelpos \ \citep[J2000, with a $1\sigma$ uncertainty of $\sim$$200$\,mas;][]{Snelders_2024_ATel}. The final correlation, of both the continuum data and the burst data of \ronefourseven, were referenced to a phase-center of \evnatelpos \ (J2000), using the same frequency resolution as the calibrator data. The continuum data have $2$\,s integrations; the burst data were coherently dedispersed, using a DM of $527.7$\,pc\,cm$^{-3}$, and correlated using various `gates' that depend on the temporal widths and arrival-times of the bursts.

The correlated data were processed and calibrated with \texttt{AIPS} \citep{AIPS}. We made use of the standard EVN calibration table, which contains parallactic angle corrections and {\it a-priori} gain correction, using the gain curves and system temperature measurements that the stations recorded during the observations. We also make use of the EVN flagging table, flagging all data from antennas which were still slewing. Additionally, we flagged $18.5$\,\% ($6$\,channels on each side) at the edges of the subbands, where antenna sensitivity significantly drops. Similarly to most FRB VLBI papers (e.g., \citealt{hewitt_2024_mnras_dec}), we made use of the \texttt{AIPS} task \texttt{VLBATECR} to correct for ionospheric dispersive delays, taking into account recent developments from \citet{Petrov_2023_AJ}.

The interferometric data of the `fringe-finder' (J1829+4844) and `phase-calibrator' (J2125+0441) were inspected and flagged for radio frequency interference (RFI) using the \texttt{AIPS} tool \texttt{spflg}, which allows the user to manually flag visibilities as a function of time, frequency and baseline. In the first step of the calibration the instrumental delay between subbands is removed using the fringe-finder scan. Next, the phase-calibrator scans are used to derive phase-solutions as a function of time and frequency for the entire observation. The phase-calibrator was imaged using \texttt{DIFMAP} \citep{difmap} to construct a model of the phase-calibrator that was used for self-calibration.

The calibration solutions were transferred to the interferometric data of the bursts. We exported the flagged and calibrated burst visibilities as \texttt{UVFITS} files with \texttt{fittp} and imported them in \texttt{CASA} with \texttt{importuvfits}. We chose this approach because the Pythonic interface of \texttt{tclean} (the imaging tool of \texttt{CASA}) facilitates the iteration over burst visibilities as a function of `scan' (in this case individual bursts) and/or antenna configuration.

\begin{figure}
    \centering
    \includegraphics[width=0.919\linewidth]{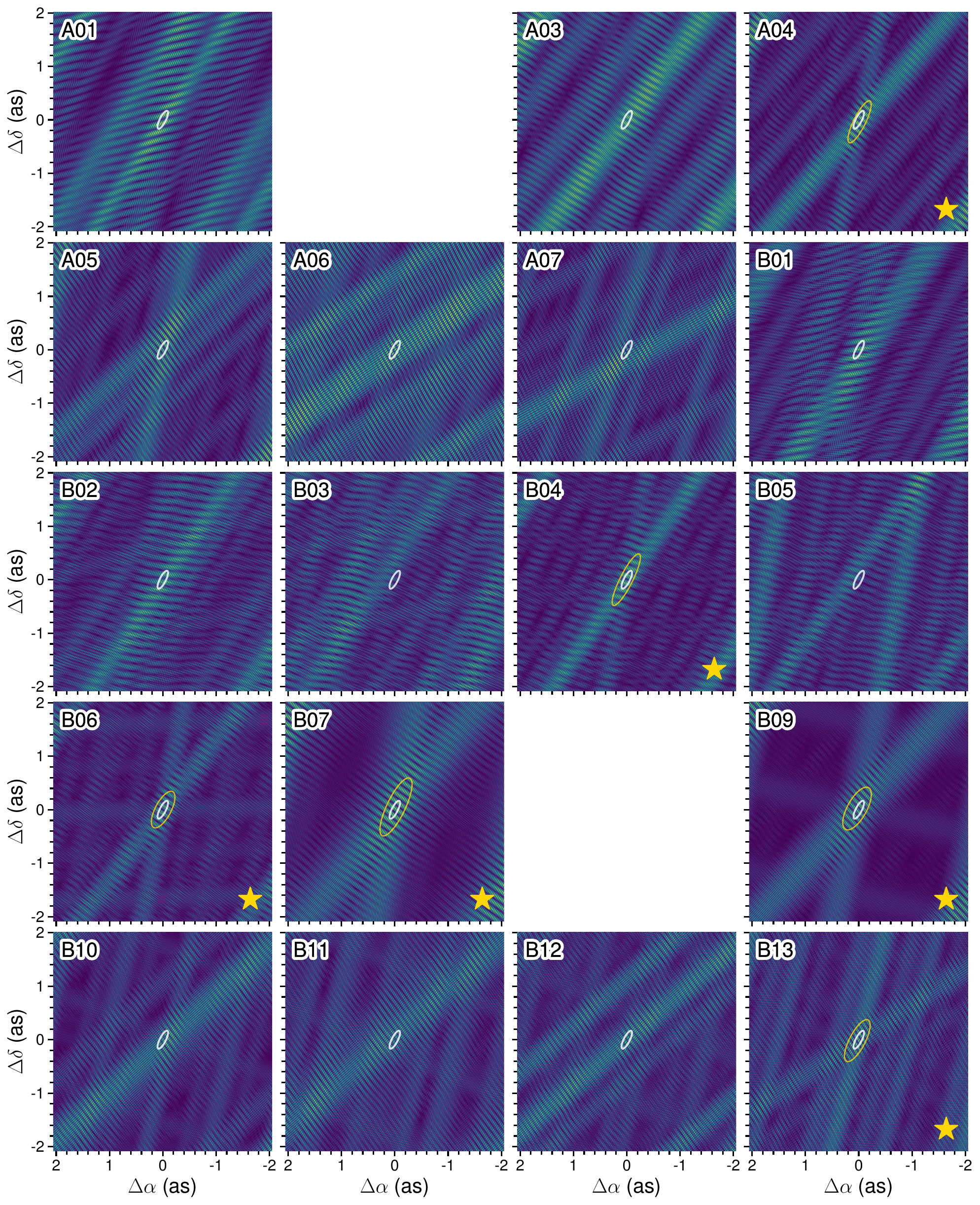}
    \caption{Each panel shows the absolute value of the dirty maps (i.e., the absolute value of the Fourier transform of the visibilities) for individual bursts. Bursts from Epoch~1 (EK056A/PR318A) are labeled A${n}$ and bursts from Epoch~2 (EK056B/PR319A) are labeled B${n}$. The bursts are sorted on their arrival times. Bursts that pass the visual inspection (as indicated by gold stars) are fitted with a rotated 2-D Gaussian function, for which the $2\sigma$ contour are shown as gold ellipses. The $2\sigma$ region of the best combined position is shown as a white ellipse in all panels. Burst A$02$ was not found to be robustly astrophysical in origin (\S\,\ref{subsubsec:burstsearch}) and no calibration solution was found for B$08$ (\S\,\ref{subsubsec:correlation}). Consequently, these panels are left blank.}
    \label{fig:burst_dirty_map}
\end{figure}

We image each burst individually using \texttt{tclean} with a natural weighting scheme, creating so-called `dirty' maps. We opt to only image using just three baselines; Effelsberg to Noto, Toru\'n and Onsala. We do this for a multitude of reasons; These are the only antennas that recorded all the bursts, recorded the widest bandwidth and since Effelsberg is so much more sensitive than the smaller antennas, including the baselines between Noto, Toru\'n and Onsala does not change the dirty maps in any meaningful way. For each burst, we image $4096 \times 4096$ pixels using a cell size of $1$\,mas and use the position quoted in \citet{Snelders_2024_ATel} as the phase-center. In Figure~\ref{fig:burst_dirty_map} we show the absolute value of the dirty maps of each burst. Burst B$08$ is not shown since no calibration solution was found in one of the neighbouring phase-calibrator scans.

We attempted to combine the calibrated burst visibilities before imaging --- similar to, e.g., \citet{Nimmo_2022_ApJL} and \citet{Kirsten_2022_Natur}. However, this actually decreased the quality of the images. This also happened when only bursts with a clear cross fringe pattern (e.g., B$07$ and B$09$) were combined. We attribute this to imperfect calibration of ionospheric phase variations. A significant contributing factor to this was the distance between the Sun and the target: $\sim$$18\degr$ during the observations, much less than the recommended $\sim$$50\degr$ for VLBI observations at these radio frequencies ($\sim$$1.4$\,GHz). Additionally, the target elevation was rather low (below $25\degr$) at the start of both observations (Figure~\ref{fig:elevation_radio_plot}). Finally, the observations took place close to the expected maximum of Solar Cycle 25\footnote{\url{https://www.swpc.noaa.gov/news/solar-cycle-25-forecast-update}}, and during the observations the solar wind conditions were changing rapidly\footnote{\url{https://www.swpc.noaa.gov/products/real-time-solar-wind}}.

To combine multiple bursts `incoherently', we first manually identify bursts that individually have a clear cross fringe pattern: A$04$ for Epoch~1 and B$04$, B$06$, B$07$, B$09$ and B$13$ for Epoch~2. For each of these bursts we fit a rotated two dimensional Gaussian function to all pixels where the absolute value of the dirty map of the burst is above $3\sigma$ \citep[as in][]{Nimmo_2022_ApJL,hewitt_2024_mnras_dec}. The $2\sigma$ uncertainty regions of those fits are shown as gold ellipses in Figure~\ref{fig:burst_dirty_map}. To combine the Gaussians we multiply the individual probabilities per pixel and fit the resulting probability with a rotated two-dimensional Gaussian.\\

Our best position for \ronefourseven is: \evnbestpos.\\ 

The $1\sigma$ uncertainty of that position is described by a two dimensional Gaussian with a major axis of $93$\,mas and minor axis of $28$\,mas that has been rotated by $27\degr$ (Figure~\ref{fig:burst_dirty_map}). The $2\sigma$ uncertainty region of the best position is shown as a white ellipse in Figure~\ref{fig:burst_dirty_map}. For almost all bursts the ellipse is on top of a fringe (e.g., A$03$) further illustrating the robustness of our localization. The angular separation between the VLBA C-band PRS detection \citep{bruni_2025_aa} and our best position is $13$\,mas, well within our $1\sigma$ uncertainty region.

We do not search for a PRS in the continuum data since the interferometric burst data showed that calibration solutions were imperfect throughout the observations --- likely due to a rapidly changing ionosphere. Even with a hypothetical perfect calibration the expected thermal noise in the continuum data would be $\sim$$20$\,$\upmu$Jy\,beam$^{-1}$ ($400$\,minute integration time with the Effelsberg, Noto, Toru\'n and Onsala telescopes) --- which is not sensitive enough to statistically claim the existence of a PRS based on the flux densities quoted in \citet{bruni_2025_aa}, \citet{bhusare_arxiv_2024} and \citet{zhang_atel_2024}.

\subsubsection{Burst Properties}\label{subsubsec:burst_properties}
\begin{figure}
    \centering
    \includegraphics[width=0.94\linewidth]{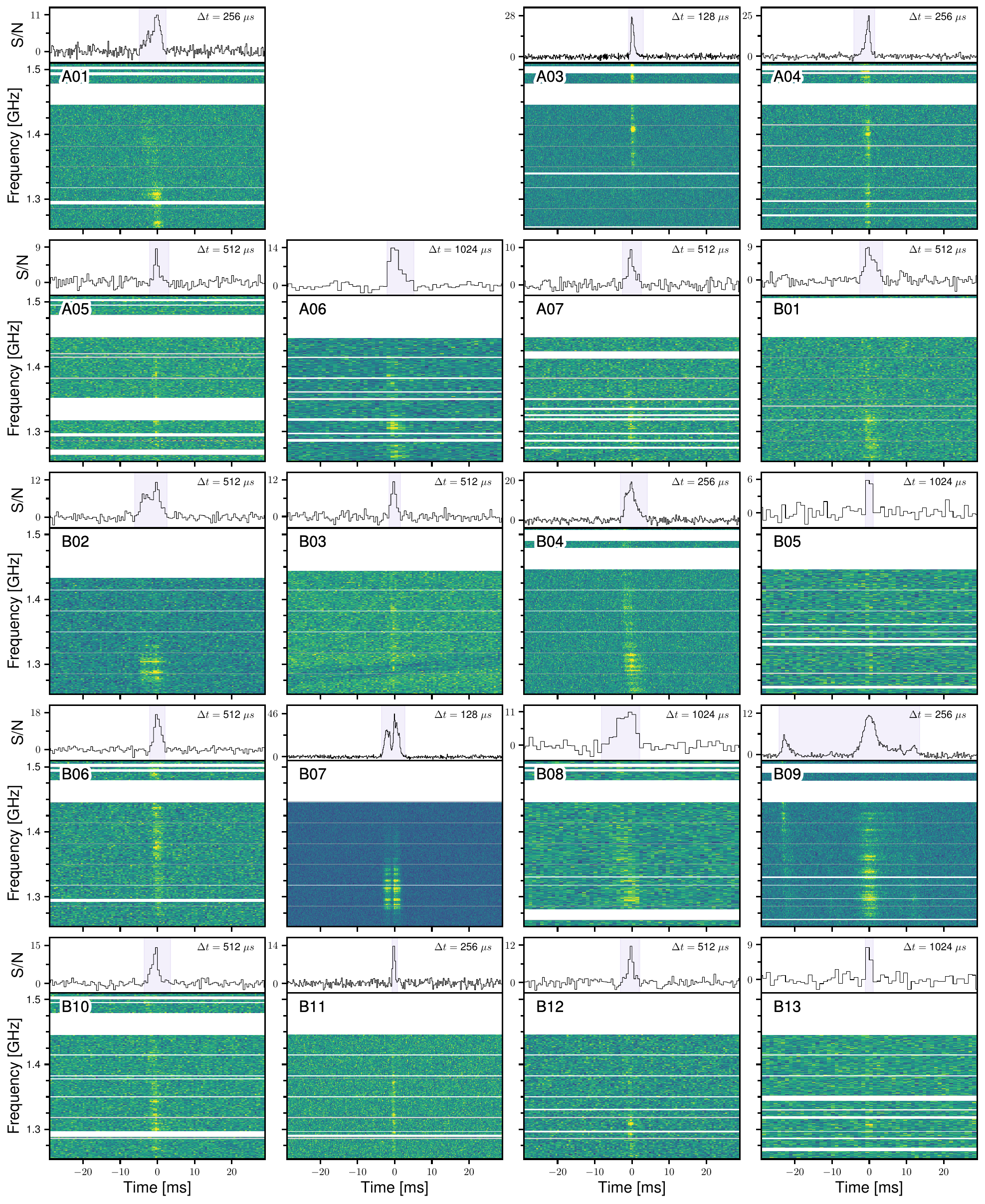}
    \caption{Dynamic spectra (bottom panels) and frequency-averaged temporal profiles (top panels) for all bursts. Each burst is coherently dedispersed to a value of $527.723$\,pc\,cm$^{-3}$. The frequency resolution in each dynamic spectrum is $500$\,kHz and the respective time resolutions are indicated in the top right of each panel. Horizontal white bands in the dynamic spectra indicate data that have been masked due to RFI or subband edges. Burst labels are the same as in Figure~\ref{fig:burst_dirty_map}. Burst A$02$ was not found to be robustly astrophysical in origin (\S\,\ref{subsubsec:burstsearch}). Consequently, this panel is left blank.}
    \label{fig:burst_family_plot}
\end{figure}

We use \texttt{DM\_$\text{phase}$} \citep{Seymour_2019_ascl} to determine the DM of burst B$07$ (the brightest burst in our sample) and find a DM of $527.723\pm0.042$\,pc\,cm$^{-3}$. We then use \texttt{SFXC} to coherently dedisperse the baseband data of just the Effelsberg telescope to create Stokes I (total intensity) files for every burst. These files have a time and frequency resolution of $32$\,$\upmu$s and $500$\,kHz, respectively, and are written out as \texttt{sigproc} filterbank files with $32$-bit floating-point numbers. The dynamic spectra of the bursts are shown in Figure~\ref{fig:burst_family_plot} where they are further averaged in time to increase their visibility. For every burst we determine their time-of-arrival, width, signal-to-noise ratio (S/N), fluence, spectral luminosity and peak flux density and tabulate the results in Table~\ref{tab:burst_properties}. 

\begin{table}[ht]
    \centering
    \caption{Burst properties.}
    \begin{tabular}{cccccccc}
     \hline
     \hline
       Burst label & ToA$^{\text{a}}$ & Det. S/N$^{\text{b}}$ & Peak S/N$^{\text{c}}$ & Width$^{\text{d}}$ & Fluence$^{\text{e}}$ & Peak Flux Density$^{\text{e}}$ & Spectral Luminosity$^{\text{e,f}}$ \\
        & [MJD] &  &  & [ms] & [Jy\,ms] & [Jy] & [$10^{31}$\,erg\,s$^{-1}$ Hz$^{-1}$] \\  
       \hline
A01 & 60355.2973790745 & 29.7 & 10.9 &  7.168 & 1.34 & 0.44 & 6.60 \\
A02$^{\text{g}}$  &    &  7.3 &      &        &      &      &      \\
A03 & 60355.3652337659 & 30.2 & 27.1 &  3.968 & 1.52 & 1.54 & 13.60 \\
A04 & 60355.3940163658 & 32.2 & 24.9 &  5.632 & 1.44 & 1.01 &  9.05 \\
A05 & 60355.4368167038 & 11.0 &  8.6 &  5.120 & 0.37 & 0.27 &  2.59 \\
A06 & 60355.4665637667 & 16.8 & 13.9 &  7.168 & 1.11 & 0.31 &  5.50 \\
A07 & 60355.4823046109 &  9.5 &  9.4 &  5.120 & 0.60 & 0.31 &  4.12 \\
B01 & 60360.2914847421 & 17.4 &  8.8 &  6.144 & 0.79 & 0.26 &  4.54 \\
B02 & 60360.3168374267 & 20.1 & 11.3 &  8.704 & 1.35 & 0.34 &  5.50 \\
B03 & 60360.3262287994 &  9.5 & 11.4 &  3.072 & 0.52 & 0.34 &  5.97 \\
B04 & 60360.3290625037 & 44.6 & 19.3 &  7.168 & 2.31 & 0.78 & 11.39 \\
B05 & 60360.3332507559 & 10.4 &  5.8 &  2.048 & 0.24 & 0.13 &  4.23 \\
B06 & 60360.3567331963 & 26.2 & 17.0 &  4.096 & 1.01 & 0.48 &  8.74 \\
B07 & 60360.3701272696 &  7.6 & 45.8 &  6.400 & 6.67 & 2.71 & 36.87 \\
B08 & 60360.3754591768 & 23.0 & 10.9 & 10.240 & 1.23 & 0.23 &  4.24 \\
B09 & 60360.3965590087 & 90.9 & 11.4 & 37.888 & 2.66 & 0.46 &  2.49 \\
B10 & 60360.4223725478 & 15.2 & 14.1 &  7.168 & 0.84 & 0.41 &  4.13 \\
B11 & 60360.4279151582 & 19.9 & 13.7 &  1.536 & 0.47 & 0.59 & 10.88 \\
B12 & 60360.4282615629 & 13.7 & 11.7 &  5.120 & 0.56 & 0.35 &  3.85 \\
B13 & 60360.4549215993 & 18.5 &  8.2 &  2.048 & 0.30 & 0.18 &  5.23 \\
       \hline
       \multicolumn{8}{l}{$^{\text{a}}$ Corrected to the Solar System Barycenter to infinite frequency assuming a dispersion measure of $527.723$\,pc\,cm$^{-3}$,} \\
       \multicolumn{8}{l}{\quad a reference frequency of $1494$\,MHz, a dispersion measure constant of $1/(2.41 \times10^{-4})$\,MHz$^{2}$\,pc$^{-1}$\,cm$^{3}$\,s and a}  \\
       \multicolumn{8}{l}{\quad source of position of $\alpha = 21^{\mathrm{h}}27^{\mathrm{m}}39\fs835$, $\delta = +04\degr19\arcmin45\farcs634$ (J2000, ICRF).}  \\
       \multicolumn{8}{l}{\quad The times quoted are dynamical times (TDB).} \\
       \multicolumn{8}{l}{$^{\text{b}}$ The detection signal-to-noise (S/N), as reported by \texttt{Heimdall}.} \\
       \multicolumn{8}{l}{$^{\text{c}}$ The peak value of the time-series, as shown in Figure~\ref{fig:burst_family_plot}.} \\
       \multicolumn{8}{l}{$^{\text{d}}$ Manually determined time-span of the burst, shown as the highlighted region in the time-series in Figure~\ref{fig:burst_family_plot}.} \\
       \multicolumn{8}{l}{$^{\text{e}}$ We estimate a (conservative) error of 20\% for these values, which is dominated by the uncertainty in the system} \\
       \multicolumn{8}{l}{\quad equivalent flux density (SEFD) of the Effelsberg telescope.} \\
       \multicolumn{8}{l}{$^{\text{f}}$ Using Equation~5 from \citet{ouldboukattine_2024_arxiv} and assuming a luminosity distance of $616$\,Mpc.} \\
       \multicolumn{8}{l}{$^{\text{g}}$ Burst A02 was removed from the sample (\S\,\ref{subsubsec:burstsearch}).} \\
    \end{tabular}
    \label{tab:burst_properties}
\end{table}

\subsection{FRB DM Budget}
\label{sec:intervening_dm}

As noted in \S\,\ref{subsubsec:burst_properties}, \ronefourseven\ has an observed dispersion measure, $\mathrm{DM}_{\mathrm{obs}}$, of $527.723 \pm 0.042~\mathrm{pc\,cm^{-3}}$. This value is unexpectedly large given the modest Galactic foreground and the low redshift of the host galaxy, as highlighted by \cite{Tian_2024_MNRAS} and \cite{chen_ApJ_2025}. To explain this excess, we decompose the total DM into five astrophysical components, all evaluated in the observer frame:

\begin{equation}
\mathrm{DM}_{\mathrm{obs}}
= \mathrm{DM}_{\mathrm{disk}}
+ \mathrm{DM}_{\mathrm{halo}}
+ \mathrm{DM}_{\mathrm{IGM}}
+ \mathrm{DM}_{\mathrm{fhalo}}
+ \mathrm{DM}_{\mathrm{host}} .
\label{eq:DM_budget}
\end{equation}

Here, $\mathrm{DM}_{\mathrm{disk}}$ and $\mathrm{DM}_{\mathrm{halo}}$ represent the Milky Way disk and halo contributions, respectively. $\mathrm{DM}_{\mathrm{IGM}}$ arises from the intergalactic medium, $\mathrm{DM}_{\mathrm{fhalo}}$ reflects excess dispersion from intervening galaxy, group, or cluster halos along the line of sight, and $\mathrm{DM}_{\mathrm{host}}$ comprises the host galaxy's interstellar medium and circumgalactic halo. Below, we quantify the DM contribution of each component, specify the adopted priors, and propagate their uncertainties to identify the dominant contributor to the total dispersion.

\subsubsection{Milky Way Disk and Halo Contributions}
\label{subsec:MW_DM}

For $\mathrm{DM}_{\mathrm{disk}}$, we employ a Gaussian prior $\mathcal{N}\!\bigl(\upmu_{\mathrm{disk}},\sigma_{\mathrm{disk}}^{2}\bigr)$, where $\upmu_{\mathrm{disk}} = 45~\mathrm{pc\,cm^{-3}}$ is the mean of the values predicted by the NE2001 \citep[$50$\,pc\,cm$^{-3}$;][]{cordes2002ne2001} and YMW16 \citep[$40$\,pc\,cm$^{-3}$;][]{yao2017new} Galactic free-electron-density models. We adopt a conservative fractional uncertainty of $20$\,\%, yielding $\sigma_{\mathrm{disk}} = 9\,\mathrm{pc\,cm^{-3}}$. For the Galactic halo contribution, we follow the analytic prescription of \citet{yamasaki2020galactic}. In line with previous work \citep[e.g.,][]{2025arXiv250302947A}, we use a uniform prior $\mathrm{DM}_{\mathrm{halo}} \sim \mathcal{U}\!\bigl[22.5,\,67.5\bigr]~\mathrm{pc\,cm^{-3}}$, which corresponds to a $\pm50$\,\% systematic uncertainty around the model prediction of $\mathrm{DM}_{\mathrm{halo}} = 45\,\mathrm{pc\,cm^{-3}}$ along the FRB line of sight.

\subsubsection{Intergalactic Medium Contribution}
\label{subsec:IGM_DM}

We estimate the DM from the diffuse IGM using the empirical redshift-DM relation \citep{Macquart_2020_Natur}. For the host-galaxy redshift at $z = 0.130287$ (\S\,\ref{subsec:optical_analysis}), this formalism yields $\mathrm{DM}_{\mathrm{IGM}} = 95^{+55}_{-20}$\,pc\,cm$^{-3}$ ($68$\,\% credible region) with a mean $\mathrm{DM}_{\mathrm{IGM}} = 116$\,pc\,cm$^{-3}$ and a standard deviation of $\mathrm{DM}_{\mathrm{IGM}} = 61$\,pc\,cm$^{-3}$. To construct the associated probability-density function, we follow the analytic prescription of \cite{Macquart_2020_Natur}, adopting $\alpha = 3.0$ and $\beta = 3.0$ for the power-law indices of the inner-halo baryon profile, and a fluctuation parameter $F = 0.2$. These parameter choices are consistent with recent FRB studies \citep[e.g.,][]{2024ApJ...965...57B}.\\

\subsubsection{Intervening Halo Contributions}
\label{subsec:foregrounfdm}
Here we estimate the DM contribution from intervening structures, such as foreground galaxies, groups, or clusters, along the FRB line of sight. \cite{OConnor_2024_ATel} utilized CHIME/FRB baseband localization \citep{Shin_2024_ATel} and the DESI Legacy Imaging Survey-based cluster catalog of \cite{WenHan2024} to identify a candidate foreground galaxy cluster, J212719.9+042225. Subsequent analysis by \cite{chen_ApJ_2025} confirmed that this cluster likely contributes significantly to the FRB's DM budget. The cluster, located at a spectroscopic redshift $z=0.0903$ \citep{WenHan2024}, lies $5.6\arcmin$ from the FRB EVN position, well within its virial radius. Therefore, we expect a non-negligible DM contribution from its intracluster medium (ICM).

This cluster has a derived total mass of $M_{500} = 6.8 \times 10^{13}\,M_{\odot}$ and a radius of $r_{500} = 0.67\,\mathrm{Mpc}$, based on DESI Legacy Imaging Survey data \citep{WenHan2024}. Here, $M_{500}$ represents the total mass enclosed within a sphere of radius $r_{500}$, where the average matter density is $500$ times the critical density of the Universe at the cluster's redshift.

To quantify the DM contribution from the ICM of J212719.9+042225, we model the electron density distribution using a spherical $\beta$-model:
\begin{equation}
n_e(r) = n_{e0}\left[1+\left(\frac{r}{r_c}\right)^2\right]^{-3\beta/2},
\end{equation}
where $n_{e0}$ is the central electron density, $r_c$ is the core radius, and $\beta$ is the slope parameter. We adopt $\beta = 0.67$, consistent with \citet{Cavaliere1976} and \citet{Sarazin1986}. The core radius is assumed to be $r_c = 0.1\,r_{500}$, following common practice in X-ray studies of galaxy clusters \citep[e.g.,][]{Mohr1999,Vikhlinin2006}. The normalization $n_{e0}$ is derived by imposing that the integrated gas mass within $r_{500}$ equals a fraction $f_{\rm gas}$ of the total cluster mass:
\begin{equation}
f_{\rm gas}\,M_{500} = 4\pi\,\upmu_e\,m_p\int_0^{r_{500}} n_e(r)\,r^2\,dr,
\end{equation}
where we adopt $f_{\rm gas} = 0.12$ \citep{Allen2004,Vikhlinin2006}, $\upmu_e \approx 1.167$ is the reduced mass after accounting for fully ionized helium and hydrogen, and $m_p$ is the proton mass.

Using this electron density model and converting the FRB's angular offset into a projected distance of $588$\,kpc, our numerical integration yields an estimated ICM DM of $\mathrm{DM}_{\mathrm{ICM}} \approx 66$\,pc\,cm$^{-3}$. Alternatively, in the observer frame, $\mathrm{DM}_{\rm ICM} \approx 60$\,pc\,cm$^{-3}$.

Next, we searched for additional foreground groups. No such groups were found in the SDSS~DR12 and 2MRS catalogs \citep{2016A&A...596A..14S} whose reported group radii were larger than their projected angular separation from the FRB position at the cluster redshift. Note that while \citet{chen_ApJ_2025} identified a dominant DM contribution from a foreground cluster in the case of FRB~20220610A, their analysis did not explicitly model the DM from other intervening galaxy halos. Here, we extend this approach by also estimating the cumulative DM contribution from all foreground galaxy halos along the sightline. We queried the DESI Legacy Imaging Survey photometric redshift catalog \citep{2019ApJS..242....8Z} for galaxies within $2\degr$ of the FRB position. We selected galaxies satisfying $z_{\rm photo} - 2\sigma_z < 0.130287$, where $0.130287$ is the host redshift (\S\,\ref{subsec:optical_analysis}). For each candidate, we computed the ratio of its effective (half-light) radius to the projected angular offset from the FRB, hereafter referred to as the `offset ratio'. We selected galaxies with offset ratios exceeding the empirically measured stellar-to-halo radius ratio (SRHR) of $0.018$ at $z\sim0.1$ \citep{Somerville2018}, yielding eight candidates whose virial radii overlap with the FRB sight line (Appendix~\ref{appendix:intersection}). Their individual DM contributions are estimated in Appendix~\ref{appendix:intersection}, yielding a total DM contribution from the foreground galaxy halos of $\mathrm{DM}_{\rm c,halo} = 135 \pm 68$\,pc\,cm$^{-3}$.

Hence, the total intervening-halo contribution is:
\[
\mathrm{DM}_{\rm fhalo} = \mathrm{DM}_{\rm c,halo} + \mathrm{DM}_{\rm ICM} = 60 + (135 \pm 68) \,\mathrm{pc\,cm}^{-3}.
\]

Among the eight foreground galaxies, one galaxy, DESI J212740.488+041911.64 (hereafter `the central galaxy'), is located just $35.5\arcsec$ from the FRB host and is classified as a spiral based on its morphology in DESI images \citep{2025A&A...694A.271C}. Given its sky proximity to the FRB host and their small photometric redshift difference, the central galaxy may share the same dark-matter halo as the FRB host. We analyze the spectroscopic properties of this central galaxy and its relationship to the FRB host in detail in \S\,\ref{subsec:optical_analysis} and further discuss their potential shared halo in \S\,\ref{sec:central_galaxy}.

\subsubsection{Host Galaxy Contribution}
\label{subsec:hostdm}

We estimate the contribution of the \ronefourseven\ host galaxy to the observed DM using both emission-based diagnostics \citep[see, for example,][]{2025AA...696A..81B} and a DM budget subtraction.

We begin by modeling the host galaxy's surface brightness profile. A S\'ersic profile fit with index $n=1$ to the DESI Legacy Survey $r$-band image yields an effective angular radius of $a = 0.4\arcsec$. Within a circular aperture of this radius, we measure an extinction-corrected H$\alpha$ flux of $F = \left( 1.54^{+0.10}_{-0.08} \right) \times10^{-16}$\,erg\,s$^{-1}$\,cm$^{-2}$. Dividing by the aperture area, $\pi a^2 \simeq 0.503$\,arcsec$^{2}$, gives an observed surface brightness of $S'_{\rm obs} \simeq 54.1^{+3.5}_{-2.8}$\,Rayleigh. After correcting for cosmological surface-brightness dimming via $(1+z)^4$ with $z = 0.130287$ (\S\,\ref{subsec:optical_analysis}), we obtain a source-frame surface brightness of $S \left( \mathrm{H} \alpha \right)_\mathrm{source} \simeq 88.1^{+5.7}_{-4.6}\,\mathrm{Rayleigh}$.

This value is converted to an emission measure (EM) using the relation appropriate for a warm ionized medium at $T = 10^4\,\mathrm{K}$ \citep{1977ApJ...216..433R}:
\[
\mathrm{EM} = 2.75\,\mathrm{pc\,cm}^{-6} \left( \frac{T}{10^4\,\mathrm{K}} \right)^{0.9} \left[ \frac{S \left( \mathrm{H} \alpha \right)_\mathrm{source}}{\mathrm{Rayleigh}} \right],
\]
yielding $\mathrm{EM} \simeq 242^{+16}_{-13}$\,pc\,cm$^{-6}$.

To translate EM into a DM, we adopt the model:
\[
\mathrm{DM}_{\rm host} = \sqrt{\frac{\mathrm{EM}\,L}{A}},
\]
where $L = 2a = 1.84\times10^3$\,pc is the physical path length through the emitting region, and $A = \zeta\,(1+\epsilon^2)/f$ encodes the microphysical properties of the ionized gas. Following \citet{2022ApJ...931...88C}, we impose a flat prior $A \in [1, 50]$ and marginalize to obtain $\mathrm{DM}_{\rm host} = 132^{+92}_{-29}$\,pc\,cm$^{-3}$.

In the observer frame, this corresponds to a host contribution of:
\[
\mathrm{DM}_{\rm host,\,obs} = \frac{\mathrm{DM}_{\rm host}}{1+z} = 117^{+81}_{-26}\,\mathrm{pc\,cm}^{-3}.
\]
Adding the contribution from the host's circumgalactic halo, estimated to be $9$\,pc\,cm$^{-3}$ in the observer frame (Appendix~\ref{appendix:intersection}), we find a total host-frame contribution of $\mathrm{DM}_{\rm host,\,obs} = 126^{+81}_{-26}$\,pc\,cm$^{-3}$.

We independently verify this estimate using a DM budget analysis:
\[
\mathrm{DM}_{\rm host,\,obs} = \mathrm{DM}_{\rm obs} - \left( \mathrm{DM}_{\rm disk} + \mathrm{DM}_{\rm halo} + \mathrm{DM}_{\rm IGM} + \mathrm{DM}_{\rm fhalo} \right).
\]
Adopting mean values for these components from \S\,\ref{subsec:MW_DM}, \S\,\ref{subsec:IGM_DM}, and \S\,\ref{subsec:foregrounfdm} as $\{45, 45, 116, 195\}$\,pc\,cm$^{-3}$ respectively, the sum of these foreground DM terms is $\sum_i \mathrm{DM}_i = 401$\,pc\,cm$^{-3}$. The combined uncertainty is $\sigma_{\rm fore} = \sqrt{9^2 + 22.5^2 + 61^2 + 68^2} \approx 95$\,pc\,cm$^{-3}$.

Subtracting from $\mathrm{DM}_{\rm obs} = 527.72 \pm 0.04\,\mathrm{pc\,cm^{-3}}$, we obtain $\mathrm{DM}_{\rm host,\,obs} \approx 126 \pm 95\,\mathrm{pc\,cm^{-3}}$. This subtraction-based value is in good agreement with the H$\alpha$-derived estimate, lending confidence to both methods. The substantial host contribution inferred from these independent approaches suggests a dense, ionized environment local to the FRB source, which can provide meaningful constraints on the ionization state and baryonic content of its host galaxy.

This comprehensive FRB DM budget analysis provides a robust framework for understanding the total observed dispersion of \ronefourseven, highlighting the significant contributions from both intervening structures and the host galaxy itself.

\subsection{GTC Spectroscopic Analysis of the FRB Host and Central Galaxy}
\label{subsec:optical_analysis}

We used the reduced 2-D, long-slit spectra from both the 2024 and 2025 runs to extract different apertures that correspond to the spectra of the central galaxy and the FRB host. The apertures were $2.0\arcsec$ and $2.9\arcsec$ along the slit, respectively, with a width of $1.23\arcsec$, and were centered on the positions of maximum emission for each target in both runs. In the case of the 2025 observations of the FRB host we extracted spectra from two different sets of processed data (2025A \& 2025B), one for each set of arc-lamp observations taken.

Given the relatively low continuum signal in both spectra, we could not model the continuum emission using stellar population synthesis. Instead, we obtained the parameters of the detected emission lines by analyzing the emission above a low-order polynomial continuum. Line fluxes were measured by adding all flux above the best-fitting continuum within specific spectral windows (direct flux measurements). Alternatively, we also measured line fluxes by fitting an analytical function to the profile using the \texttt{lmfit} Python package \citep{2021zndo...5570790N}. For these fits, we used both single Gaussians and more complex Gauss-Hermite profiles, which account for non-zero skewness and kurtosis. Unless otherwise stated, we use either direct measurements or single-Gaussian fits, depending on the S/N or equivalent width of the line. For determining radial velocities, however, we mostly rely on single-Gaussian (1G) fits, as they are more robust against noise than Gauss-Hermite profile fits, with emphasis on the results obtained with the higher-resolution 2025 GTC observations. For measuring line fluxes and ratios we will exclusively employ the 2024 spectra as these provide a wider wavelength coverage that includes the [O{\sc ii}]$\lambda\lambda$3726,3629\AA\ and [O{\sc iii}]$\lambda\lambda$4959,5007\AA\ doublets and the H$\beta$ Balmer line. The latter line is crucial for correcting line fluxes and ratios for dust extinction and reddening, respectively.

\begin{figure}
    \centering
    \includegraphics[width=1.0\linewidth]{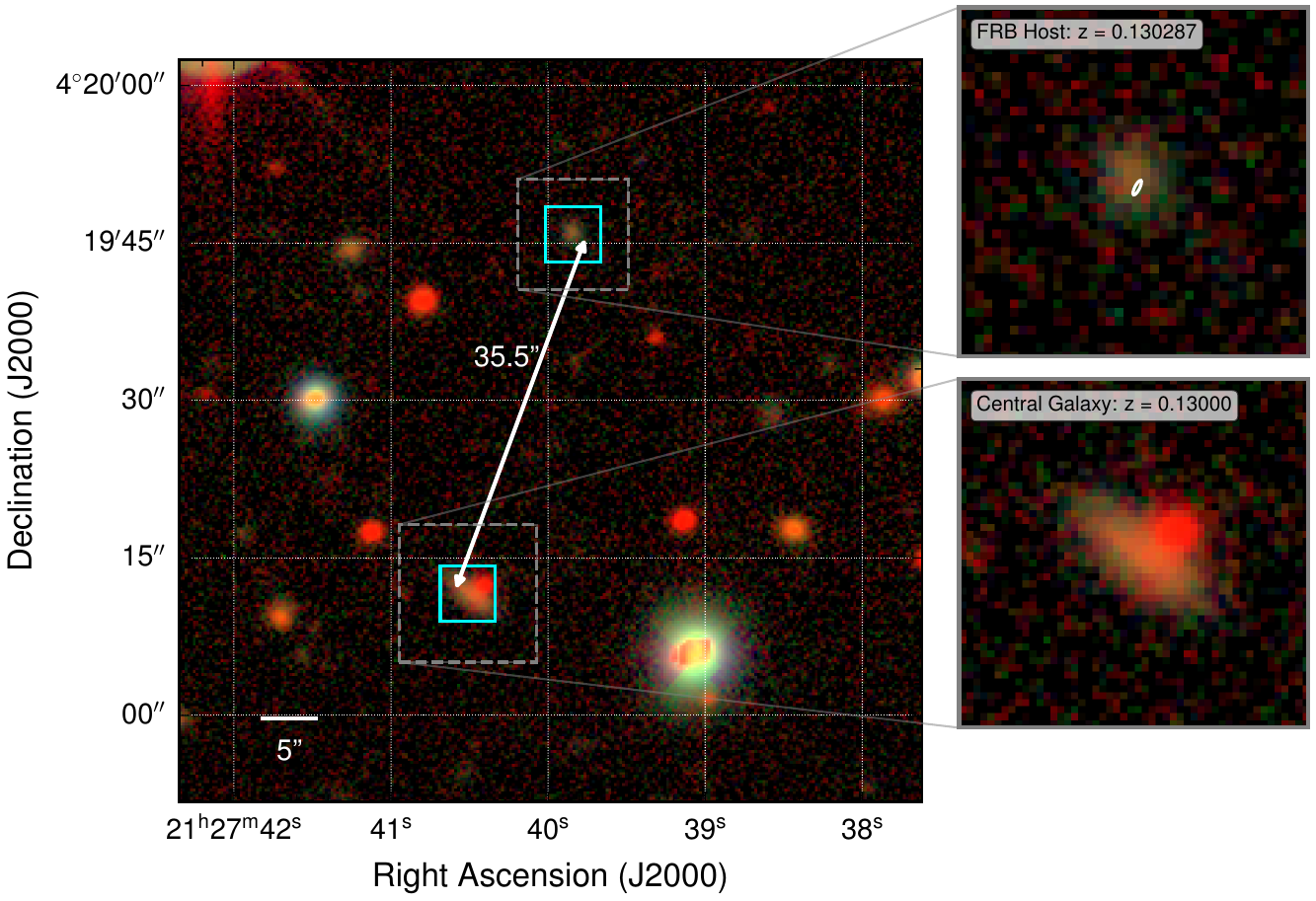}
\caption{Wide-field RGB view of the \ronefourseven field using DESI $g$ (blue), $r$ (green), and $z$ (red) band images. The FRB host galaxy at $z = 0.130287$, likely a satellite of a nearby central galaxy at the same systemic redshift, is marked with a cyan box along with its companion; their projected separation of $35.5\arcsec$ ($85$\,kpc transverse distance) is marked by the double-headed arrow. Gray dashed boxes indicate the zoom regions shown in the upper (host) and lower (central galaxy) insets. In the host inset, the EVN-derived FRB position is highlighted by a white ellipse whose semi-axes represent the $10\sigma$ localization uncertainty (inflated to be visible on this scale).}
\label{fig:R147_fov}
\end{figure}

\begin{figure}
    \centering
    \includegraphics[width=\linewidth]{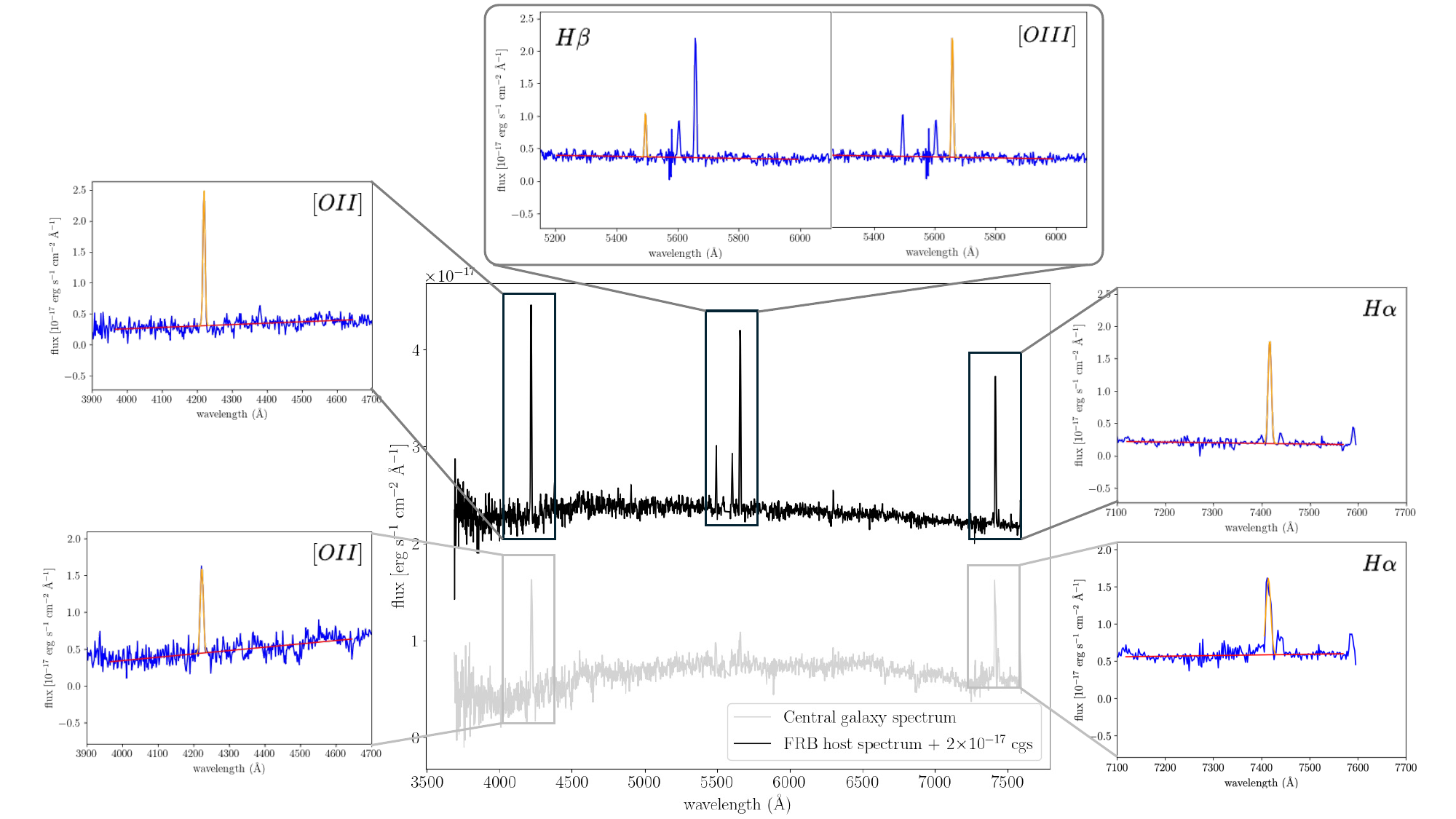}
    \caption{Central panel: OSIRIS spectra of the FRB host (black solid line) and the central galaxy (solid grey). Peripheral panels: These insets show the brightest emission lines detected in each spectrum and their corresponding single-Gaussian fits. On the left, we show the region around the unresolved [O{\sc ii}]$\lambda\lambda$3726,3629\AA\ doublet, while the H$\beta$ and [O{\sc iii}]$\lambda\lambda$4959,5007\AA\ lines and their best fits are shown on top. The right panels show the H$\alpha$ line region for each of the two spectra. All spectra correspond to our 2024 observations, as these data provide a wider wavelength coverage.}
    \label{fig:spectra_collage}
\end{figure}

Figure~\ref{fig:spectra_collage} shows the spectra of the two regions with detected lines along the slit, using the 2024 dataset and the extraction parameters listed above. We also present the results of the 1G fits for most of the emission lines detected in these spectra. The measured radial velocities from both the 2024 and 2025 runs and line fluxes from the 2024 observations are provided in Table~\ref{tab:flines}.

\begin{table}[!ht]
    \centering
    \caption{Line radial velocities and fluxes of the FRB host and the central galaxy.}
    \begin{tabular}{lccccccccccc}
    \hline
    \hline
        Line & $v_{\text{helio}}$ & $\delta v$ & $f_{\text{data}}$ & $\delta f_{\text{data}}$ & EW & $\delta$EW & $f_{1\text{G}}$ & $\delta f_{1\text{G}}$ & EW$_{1\text{G,c}}$ & $f_{\text{data,c}}$ & $f_{1\text{G,c}}$ \\
        & \multicolumn{2}{c}{[km\,s$^{-1}$]} & \multicolumn{2}{c}{[$10^{-17}$\,erg\,s$^{-1}$\,cm$^{-2}$]} & \multicolumn{2}{c}{[\AA]} & \multicolumn{2}{c}{[$10^{-17}$\,erg\,s$^{-1}$\,cm$^{-2}$]} & [\AA] & \multicolumn{2}{c}{[$10^{-17}$\,erg\,s$^{-1}$\,cm$^{-2}$]}\\ \hline
        \multicolumn{12}{c}{FRB host}\\ \hline
        {[O\sc{ii}]$_{\mathrm{2024}}$} & 39670 & 7 & 15.8 & 1.300 & 51.2 & 5.4 & 17.0 & 1.320 & 55.0 & 20.1 & 21.6 \\
        H$\beta_{\mathrm{2024}}$ & 39048 & 12 & 4.00 & 0.614 & 13.8 & 1.7 & 4.72 & 0.621 & 15.7 & 6.08 & 6.93 \\
        {[O\sc{iii}]$_{\mathrm{2024}}$} & 39013 & 4 & 14.6 & 0.753 & 40.5 & 2.4 & 15.1 & 0.758 & 41.9 & 17.2 & 17.8 \\
        H$\alpha_{\mathrm{2024}}$ & 39016 & 4 & 13.2 & 0.648 & 70.6 & 5.1 & 13.5 & 0.652 & 72.1 & 15.3 & 15.6 \\
        H$\alpha_{\mathrm{2025A}}$ & 39058.5 & 0.9 & \nodata & \nodata & \nodata & \nodata & \nodata & \nodata & \nodata & \nodata & \nodata \\
        H$\alpha_{\mathrm{2025B}}$ & 39059.7 & 0.9 & \nodata & \nodata & \nodata & \nodata & \nodata & \nodata & \nodata & \nodata & \nodata \\
        {[N\sc{ii}]$_{\mathrm{2024}}$} & 38992 & 12 & 1.05 & 0.356 & 5.5 & 1.9 & 1.09 & 0.357 & 5.8 & 1.18 & 1.23 \\
        {[N\sc{ii}]$_{\mathrm{2025A}}$} & 39053 & 4 & \nodata & \nodata & \nodata & \nodata & \nodata & \nodata & \nodata & \nodata & \nodata \\
        {[N\sc{ii}]$_{\mathrm{2025B}}$} & 39058 & 4 & \nodata & \nodata & \nodata & \nodata & \nodata & \nodata & \nodata & \nodata & \nodata \\
        {[S\sc{ii}]$_{\mathrm{2024}}$} & 39017 & 11 & 1.45 & 0.421 & 8.3 & 2.5 & 1.65 & 0.425 & 9.4 & 1.62 & 1.85 \\ \hline
        \multicolumn{12}{c}{Central galaxy}\\ \hline
        {[O\sc{ii}]$_{\mathrm{2024}}$} & 39941 & 15 & 12.2 & 1.460 & 27.4 & 3.6 & 12.5 & 1.460 & 28.1 & 45.3 & 16.1 \\
        H$\beta_{\mathrm{2024}}$ & 38858 & 48 & 1.07 & 0.730 & 4.4 & 1.0 & 2.26 & 0.738 & 6.0 & 8.87 & 5.41 \\
        {[O\sc{iii}]$_{\mathrm{2024}}$} & 38932 & 39 & 3.57 & 0.789 & 4.9 & 1.1 & 3.22 & 0.787 & 4.4 & 9.23 & 3.85 \\
        H$\alpha_{\mathrm{2024}}$ & 38887 & 23 & 12.1 & 0.745 & 22.6 & 1.4 & 12.5 & 0.748 & 23.2 & 26.5 & 15.5 \\
        H$\alpha_{\mathrm{2025C}}$ & 38971 & 2 & \nodata & \nodata & \nodata & \nodata & \nodata & \nodata & \nodata & \nodata & \nodata \\
        {[N\sc{ii}]$_{\mathrm{2024}}$} & 38906 & 37 & 3.54 & 0.724 & 6.0 & 1.2 & 3.58 & 0.724 & 6.1 & 7.03 & 4.06 \\
        {[N\sc{ii}]$_{\mathrm{2025C}}$} & 38970 & 4 & \nodata & \nodata & \nodata & \nodata & \nodata & \nodata & \nodata & \nodata & \nodata \\
        {[S\sc{ii}]$_{\mathrm{2024}}$} & 38839 & 18 & 2.41 & 0.553 & 4.0 & 0.9 & 2.44 & 0.554 & 4.1 & 4.70 & 2.76 \\ \hline
    \end{tabular}
    \label{tab:flines}
\end{table}

Based on the heliocentric recession velocities given in Table~\ref{tab:flines} for the higher-resolution 2025 observations, the redshift of the FRB host obtained from the H$\alpha$ line is $z=0.130285\pm0.000003$ using the 2025A processed dataset, and $z=0.130289\pm0.000003$ with the 2025B dataset. These numbers are compatible, within the errors, with the values obtained from those same datasets using the significantly weaker [NII][N{\sc ii}]$\lambda$$6584$\AA\ line. Note that these errors include the formal fitting errors quoted in Table~\ref{tab:flines} along with the rms ($0.02$\,\AA or $0.8$\,km\,s$^{-1}$). These values were obtained using single-Gaussian fits. The difference with respect to the Gauss-Hermite best fits is well below $1$\,km\,s$^{-1}$.

In the case of the central galaxy, the redshift obtained is $z=0.129994\pm0.000006$ using H$\alpha$ alone. This value is in excellent agreement with the DESI DR1 FastSpecFit catalog \citep{2023ascl.soft08005M}, which reports $z=0.13000\pm0.00001$. We emphasize here the importance of extracting the nuclear region of the central galaxy, as it was done in the case of our 2025C dataset, since this object shows a large velocity gradient along its major axis which is due to its large mass and relatively high inclination. This alone can explain the difference of $\sim$$80$\,km\,s$^{-1}$ between our 2024 and 2025C measurements for the central galaxy recession velocity (see Table~\ref{tab:flines}).

With respect to the FRB host, previous studies estimated slightly higher values for its redshift: \cite{chen_ApJ_2025} find $z=0.1306\pm0.0002$ while \cite{bruni_2025_aa} report a value of $z=0.13056\pm0.00003$ that overlaps in this case with our 2025 measurements at a $\sim$$9\sigma$ level. Such shifts, and also the difference between our 2025A \& 2025B measurements from those obtained from our lower-spectral-resolution 2024 observations, can be explained by sub-arcsecond slit-filling and centering errors combined with finite spectral resolution --- e.g., a $0.2\arcsec$ misalignment at $R\approx235$ yields $\sim$$180$\,km\,s$^{-1}$ --- and by uncorrected atmospheric refraction. Since our 2025A and 2025B observations were specifically designed to minimize these effects in terms of spectral resolution, slit width, airmass, etc., we adopt $z=0.130287\pm0.000002_{\mathrm{stat}}\pm0.000002_{\mathrm{sys}}$ as the systemic redshift of the FRB host in all our analyses.

The line fluxes have been first corrected for foreground Milky Way attenuation, assuming a Galactic color-excess of $\mathrm{E} \left( B-V \right) = 0.06$\,mag and the \citet{Cardelli_1989_ApJ} dust-attenuation law. We then corrected the Balmer-line fluxes for underlying stellar absorption, assuming equivalent widths in absorption for H$\beta$ and H$\alpha$ of $3$\,\AA and $2$\,\AA, respectively. Finally, the different fluxes were corrected for internal extinction using the Balmer decrement after adopting an intrinsic $f_{\mathrm{H} \alpha} / f_{\mathrm{H} \beta} = 2.86$. In the case of the FRB host, which is less luminous and likely less metal-rich than the central galaxy, this line ratio is close to the theoretical value for no extinction in some cases. The dust attenuation derived from the Balmer decrement for both galaxies agrees well with the values inferred from our stellar population synthesis. These spectroscopic results, along with our photometric data, form the foundation for characterizing the intrinsic physical properties of the FRB host and the central galaxy, which we discuss in detail in \S\,\ref{section:host_properties}.

\subsection{FRB Host and Central Galaxy Properties}
\label{section:host_properties}
We characterize the physical properties of the FRB host galaxy and the nearby central galaxy using two complementary approaches: (1) nebular emission line diagnostics based on GTC/OSIRIS spectroscopy, and (2) stellar population synthesis modeling via spectral energy distribution (SED) fitting with \texttt{Prospector}. The methodology and detailed derivations of line ratios, H$\alpha$-based star formation rates (SFRs), gas-phase metallicities, and Baldwin-Phillips-Terlevich (BPT) classifications are described in Appendix~\ref{sec:line_estimates}, while the SED modeling is detailed in Appendix~\ref{sec:appendix_prospector}. The key derived properties of the \ronefourseven\ host galaxy and the central galaxy are summarized in Table~\ref{tab:A-galaxy-properties}. We also perform a BPT diagnostic analysis (see Appendix~\ref{sec:line_estimates}) to assess nuclear activity in both galaxies, which reveals that both galaxies are actively star-forming with no signatures of AGN activity.

\begin{table}[t]
\centering
\caption{Major observables of the FRB 20240114A host and its central galaxy.}
\label{tab:A-galaxy-properties}
\small
\begin{tabular}{lcc}
\hline\hline

Parameter & FRB Host & Central Galaxy \\

\hline

R.A. (J2000) & $21^{\mathrm{h}}27^{\mathrm{m}}39\fs84$ & $21^{\mathrm{h}}27^{\mathrm{m}}40\fs49$ \\

Dec. (J2000) & $+04\degr19\arcmin45\farcs8$ & $+04\degr19\arcmin11\farcs6$ \\

Galaxy name & DESI J212739.84+041945.8 & DESI J212740.49+041911.6 \\

Apparent $r$-band mag ($m_r$, AB)$^a$ & 21.94 & 20.03 \\

Galactic $r$-band extinction ($A_r$, AB)$^a$ & 0.16 & 0.16 \\

Spectroscopic redshift ($z_{\rm spec}$) & $z=0.130287\pm0.000002_{\mathrm{stat}}\pm0.000002_{\mathrm{sys}}$$^{d}$ & $0.13000\pm0.00001$$^{e}$ \\

Absolute $r$-band mag ($M_r$, AB) & $-17.22$ & $-18.97$ \\

Effective radius, $R_{\rm eff}$ (kpc)$^a$ & 0.9 & 4.3 \\

SFR (H$_\alpha$; $M_\odot$\,yr$^{-1}$)$^b$ & $0.061^{+0.004}_{-0.003}$ & $0.16^{+0.07}_{-0.01}$ \\

SFR (0-100 Myr; $M_\odot$\,yr$^{-1}$)$^c$ & $0.06_{-0.04}^{+0.06}$ & $0.52^{+0.70}_{-0.27}$ \\

Stellar mass $\log(M/M_\odot)$ $^c$ & $8.55^{+0.12}_{-0.14}$ & $9.83^{+0.09}_{-0.10}$$^c$ \\

Stellar metallicity $\log(Z/Z_\odot)$ $^c$ & $-1.24^{+0.36}_{-0.52}$ & $-1.01^{+0.33}_{-0.35}$$^c$ \\

Gas-phase metallicity $\log(Z/Z_\odot)_{\rm gas}$ $^b$ & $-0.47 \pm 0.13$ & $-0.23 \pm 0.13$ \\

Mass-weighted age (Gyr)$^c$ & $5.79_{-1.24}^{+0.94}$ & $6.03_{-1.65}^{+0.85}$ \\

$\log(\mathrm{sSFR}_{0-100\,\rm Myr})$ (yr$^{-1}$)$^c$ & $-9.76^{+0.44}_{-0.49}$ & $-10.11^{+0.47}_{-0.41}$ \\

$A_{V}$ (young stars; mag)$^c$ & $0.11^{+0.14}_{-0.08}$ & $1.12^{+0.32}_{-0.32}$ \\

Host AGN & N & N \\

\hline

\end{tabular} \\
$^a$ Using data from the DESI Legacy Survey photometric catalog \citep{2019ApJS..242....8Z}.\\

$^b$ Obtained from optical spectra (Appendix \ref{sec:line_estimates}).\\

$^c$ Estimated using \texttt{Prospector}.\\

$^{d}$ See Section~\ref{subsec:optical_analysis} for more details on the uncertainties associated to this value.\\ 

$^{e}$ From the DESI Data Release 1 (DR1) FastSpecFit Spectral Synthesis and Emission-Line Catalog \citep{2023ascl.soft08005M} \\

\end{table}

\begin{figure}[htbp]
\centering
    \begin{minipage}[b]{0.49\textwidth}
        \centering
        \includegraphics[width=\textwidth]{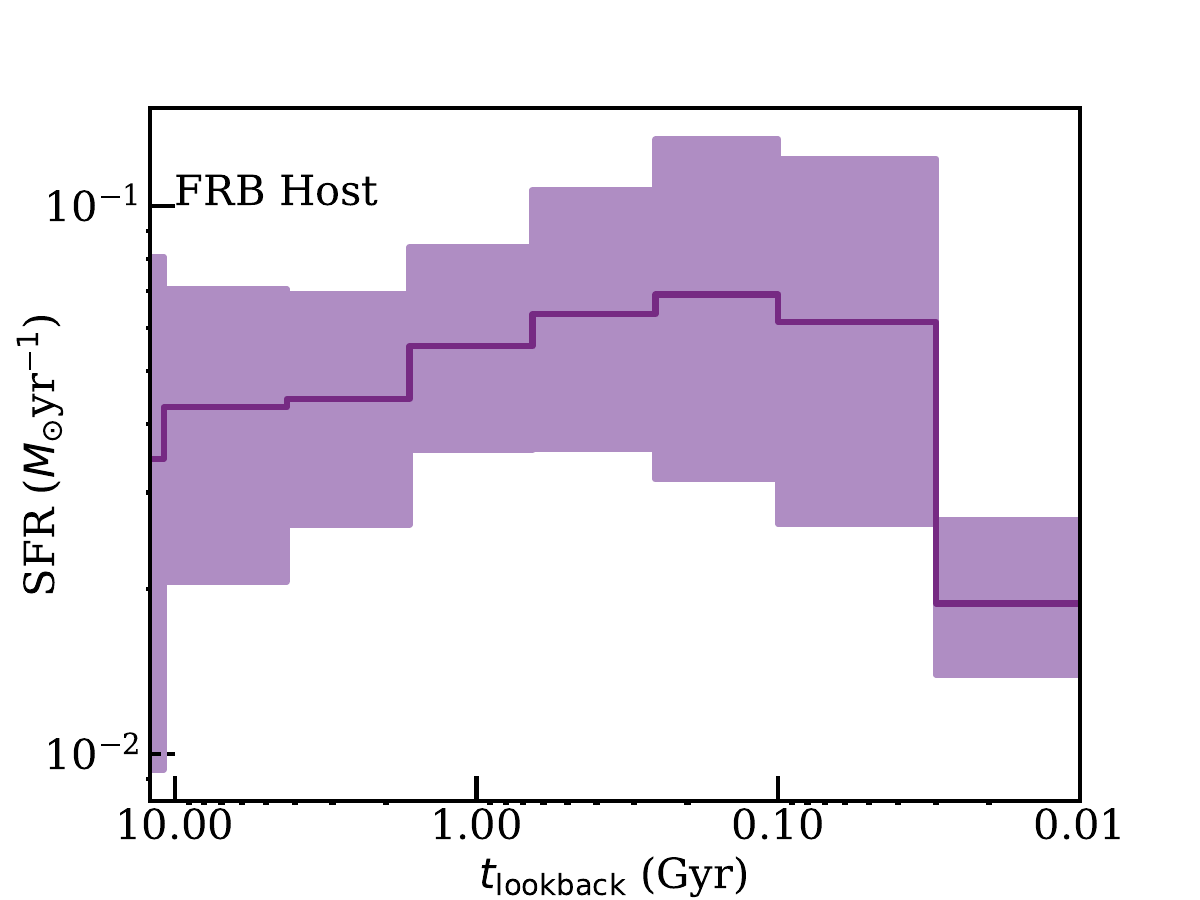}
    \end{minipage}
    \hfill
    \begin{minipage}[b]{0.49\textwidth}
        \centering
        \includegraphics[width=\textwidth]{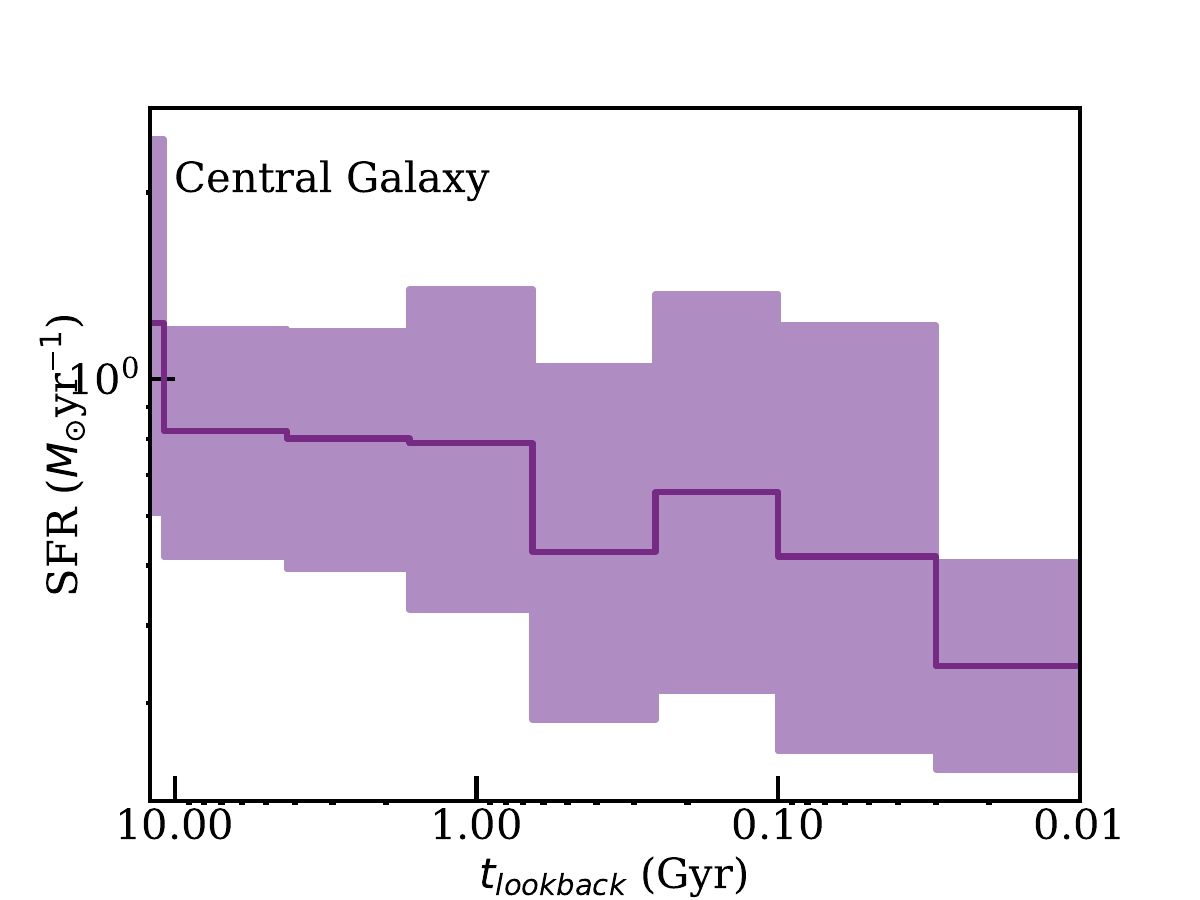}
    \end{minipage}
\caption{Nonparametric star-formation histories (SFHs). The left panel shows the SFH of the FRB host galaxy, and the right panel that of the central companion, both reconstructed with \texttt{Prospector} (see \S\,\ref{section:host_properties} and Appendix~\ref{sec:appendix_prospector}). Solid lines indicate the median SFR in each time bin; shaded envelopes mark the 16th--84th percentile credible intervals.}
\label{fig:SFH}
\end{figure}

The FRB host is a dwarf galaxy ($\log (M/M_\odot) = 8.6$) exhibiting sub-solar gas-phase metallicity, $\log(Z/Z_\odot) = -0.46 \pm 0.17$, and an elevated SFR of $0.06\,M_{\odot}$\,yr$^{-1}$, which is in excellent agreement with the dust-corrected H$\alpha$ estimate. In contrast, the central galaxy is more massive ($\log (M/M_\odot) = 9.8$), more metal-rich ($\log(Z/Z_\odot) = -0.25 \pm 0.19$), and forms stars at $\sim$$0.52\,M_{\odot}$\,yr$^{-1}$. Moreover, both galaxies are classified as star-forming based on their specific SFRs \citep{salim_2007_ajps}. We note that our derived host properties are broadly consistent with \cite{chen_ApJ_2025}. However, our SFR is approximately $6$ times smaller than what \cite{bruni_2025_aa} estimated, who report an extinction-corrected SFR of $0.36\,M_\odot$\,yr$^{-1}$ driven by an anomalously high Balmer decrement ($H\alpha / H\beta \simeq 8$). We attribute their elevated ratio to differential slit losses and atmospheric refraction, as both our data and that of \cite{chen_ApJ_2025} --- which used wider slits --- yield $H\alpha / H\beta \sim 2.9$, indicating minimal nebular extinction.

Our nonparametric star-formation histories (SFHs; Figure~\ref{fig:SFH}) further reveal that the FRB host experienced a burst of star formation $\approx 30$--$50$\,Myr ago, analogous to those seen in the Magellanic Clouds \citep{2009AJ....138.1243H}. In contrast, the central galaxy displays a smoothly declining SFR, akin to the Milky Way's evolutionary history \citep{2022MNRAS.513L..40M}. Together, these analyses paint a coherent picture of two relatively young, gas-rich systems. Building on these individual characterizations, we now assess the evidence for their association as a satellite-central system in \S\,\ref{sec:central_galaxy}.

\subsection{Assessing the Satellite Association of the FRB Host}
\label{sec:central_galaxy}
In this section, we evaluate the prospects of the FRB host and the central galaxy forming a satellite-central system. The angular separation between the two galaxies ($35.5\arcsec$) translates to a projected galactocentric distance $R_{\rm proj} \approx 85$\,kpc at $z = 0.130287$.

From Table~\ref{tab:A-galaxy-properties}, the central galaxy's stellar mass is $\approx 6.8 \times 10^{9}\,M_\odot$. This implies a host halo mass $M_{\rm halo} \approx 3 \times 10^{11}\,M_\odot$, derived using the stellar-to-halo mass relation from \cite{Moster2010}. Using this halo mass, the virial radius ($R_{\rm vir}$) and virial velocity ($V_{\rm vir}$) are calculated from standard definitions. The virial radius, defined as the radius where the average halo density is $200$ times the critical density ($\rho_{\rm crit}$) of the Universe at the central galaxy's redshift, is

\begin{equation}
R_{\rm vir} = \left(\frac{3 M_{\rm halo}}{4 \pi \Delta_{\rm vir} \rho_{\rm crit}(z)}\right)^{1/3} \approx 132~\mathrm{kpc}\,,
\label{Eq:Rvir}
\end{equation}

where $\Delta_{\rm vir}=200$. The corresponding virial velocity is

\begin{equation}
V_{\rm vir} = \sqrt{\frac{G M_{\rm halo}}{R_{\rm vir}}} \approx 98~\mathrm{km\,s^{-1}}\,.
\end{equation}

The escape velocity, relevant for a bound orbit, is estimated as $v_{\rm esc} \approx \sqrt{2}\,V_{\rm vir} \approx 139$\,km\,s$^{-1}$. The FRB host and its putative central companion show the redshift offset of $\Delta z = 0.000287$ (Table~\ref{tab:A-galaxy-properties}) which we used to calculate their line-of-sight separation. This corresponds to a line-of-sight velocity offset relative to the central galaxy of $v_{||} = c\,\Delta z \approx 86$\,km\,s$^{-1}$.

The observed projected separation $R_{\rm proj} \approx 85$\,kpc is well within the central galaxy's virial radius ($R_{\rm vir} \approx 132$\,kpc), corresponding to $R_{\rm proj}/R_{\rm vir} \approx 0.64$. Crucially, the observed line-of-sight velocity offset ($v_{||} \approx 86$\,km\,s$^{-1}$) is considerably less than the estimated escape velocity ($v_{\rm esc} \approx 139$\,km\,s$^{-1}$). While we only observe the line-of-sight component, this indicates that the FRB host galaxy's kinematics are fully consistent with bound orbital motion within the central galaxy's gravitational potential. It is noted that residual velocity differences may stem from intrinsic peculiar motions of satellites (up to a few hundred\,km\,s$^{-1}$), measurement uncertainties in systemic redshift (typically $\sim20$--$30$\,km\,s$^{-1}$), or minor contributions from gravitational redshift or large-scale flows. These effects are comparable to the escape velocity threshold. 

Next, we examine the FRB host's projected phase-space coordinates $(R_{\rm proj}/R_{\rm vir}\approx0.64,\ v_{\parallel}/V_{\rm vir}\approx0.87)$ in the context of well-measured Local Group satellites and predictions from $\Lambda$CDM simulations. Among Milky Way satellites, Leo~I provides a close velocity analog: HST proper-motion measurements yield a Galactocentric line-of-sight velocity of $V_{\rm rad} = 167.9 \pm 2.8$\,km\,s$^{-1}$ with modest tangential motion \citep{2013ApJ...768..139S}. Relative to a fiducial Milky Way virial velocity of $V_{\rm vir} \approx 180$\,km\,s$^{-1}$ \citep{2006MNRAS.369.1688D}, this corresponds to $v_{\parallel} / V_{\rm vir} \simeq 0.93$. Leo~I resides near the virial boundary ($R \simeq 260$\,kpc); \citealp{McConnachie2012}), so while its normalized radius ($R_{\rm proj} / R_{\rm vir} \gtrsim 1$) exceeds that of the FRB host, its velocity ratio demonstrates that \(v_{\parallel}/V_{\rm vir}\sim0.9\) is readily achieved by bound satellites. On the M31 side, IC~10 illustrates comparable radii with moderately high normalized velocities: its projected separation is $\sim250$--$270$\,kpc, i.e. $R_{\rm proj} / R_{\rm vir} \sim 0.8$--$0.9$ for an M31 virial radius of $\sim$$300$\,kpc, and its relative speed is of order $150$--$220$\,km\,s$^{-1}$ depending on the adopted M31 velocity vector \citep{McConnachie2012, 2007A&A...462..101B, 2012ApJ...753....8V}. Thus, although no single Local Group satellite simultaneously matches both the FRB host's radius and velocity ratios, the combination ($R_{\rm proj} / R_{\rm vir} \sim 0.6$, $v_{\parallel} / V_{\rm vir} \sim 0.9$) lies squarely within the observed envelope spanned by well-studied satellites. High-resolution cosmological simulations further confirm that satellites at $R_{\rm proj} \sim 0.5$--$0.8$\,$R_{\rm vir}$ frequently exhibit $|v_{\parallel}| / V_{\rm vir} \sim 0.6$--$1.0$, reflecting a mixture of moderately eccentric and more radial orbits. In particular, projected phase-space libraries constructed from $N$-body simulations (e.g., MultiDark) recover broad, high-probability bands in this regime \citep{2013MNRAS.431.2307O}. Consequently, the FRB host's phase-space location is well within the statistically common satellite distribution and does not require an unusual or transient dynamical state.

Finally, we quantify the probability that the FRB host galaxy is an unrelated field galaxy (an `interloper') by chance alignment. We define a redshift-space cylinder using the central galaxy's virial properties. The cylinder's projected radius is $R_{\rm vir} = 0.132$\,Mpc. For the line-of-sight depth, we consider a window of $\pm 3\sigma_{\rm halo}$, where $\sigma_{\rm halo} \approx 0.8 \times V_{\rm vir}$. With an updated $V_{\rm vir} = 98$\,km\,s$^{-1}$, this results in $\sigma_{\rm halo} \approx 78.4$\,km\,s$^{-1}$, and a total velocity window of $2 \times 3\sigma_{\rm halo} \approx 470$\,km\,s$^{-1}$. The redshift-space depth of this cylinder is therefore $\Delta Z = (470\,\mathrm{km\,s}^{-1})/{H(z)} \approx 6.5~\mathrm{Mpc}$. The volume of this redshift-space cylinder is $V_{z} = \pi R_{\rm vir}^2 \times \Delta Z \approx 0.36$\,Mpc$^{3}$. Using a cosmic number density for dwarf galaxies with stellar masses comparable to or more massive than the FRB host ($M_{\mathrm{stellar}} = 10^{8.3} M_{\odot}$), $n_g \approx 0.0075$\,Mpc$^{-3}$ \citep{Baldry2008}, the expected number of random interlopers in this volume is $N = n_g \times V_{z} \approx 0.0027$. Assuming Poisson statistics, the probability of detecting one or more such field galaxies by chance is $P(X \ge 1) = 1 - e^{-N} \approx 0.0027$.

In summary, the FRB host galaxy's position and velocity relative to the central galaxy are kinematically consistent with a bound orbit, with its line-of-sight velocity well below the estimated escape velocity. Its phase-space location is statistically common among observed and simulated satellite populations. Furthermore, the quantitative assessment of the chance alignment hypothesis demonstrates a very low probability of the FRB host being an unrelated field galaxy. Synthesizing these lines of evidence, we conclude that the FRB host galaxy is likely to be a gravitationally bound satellite of the central galaxy.

Finally, we note that the relative properties of the two galaxies are consistent with expectations for a satellite-central system. The FRB host has a stellar mass over one-tenth that of the central galaxy, in line with typical satellite-central mass ratios in group environments \citep[][]{2013ApJ...770...57B,2015ApJ...804..136W}. It is also $\sim$$0.25$\,dex more metal-poor, consistent with the mass-metallicity relation \citep{Gallazzi2005} and with metallicity differences observed in star-forming satellite-central pairs \citep{2012MNRAS.425..273P}. Furthermore, the host's low metallicity and active star formation resemble those of the Magellanic Clouds, a well-known gas-rich companion to the Milky Way. While the analogy is illustrative, we note that the central galaxy in our system is less massive than the Milky Way, suggesting that the FRB host-central pair resides in a lower-mass group regime.

\section{Discussion and Conclusion}\label{sec:discussion_conclusion}

In this work, we present the milliarcsecond-precision EVN localization of \ronefourseven ($\sim$$90 \times 30$\,mas, $1\sigma$), representing a couple-thousand-fold improvement in sky-area uncertainty over the previous MeerKAT position \citep{Tian_2024_MNRAS}. At $z=0.130287$, this corresponds to $\lesssim100$\,pc precision, firmly placing the FRB source $0.5$\,kpc ($0.6\,R_e$) from its dwarf host nucleus. Although ground-based imaging cannot yet resolve its immediate environment, space-based instruments such as \textit{Hubble Space Telescope} and \textit{James Webb Space Telescope}, or high-resolution ALMA observations, could search for coincident star clusters or giant molecular clouds, thereby discriminating between young magnetar and compact merger progenitors. We also confirm positional agreement (within $13$\,mas) with the PRS candidate presented by \cite{bruni_2025_aa}, but do not comment further on the nature of the PRS using our EVN data because they do not reach a sufficient depth or calibration accuracy.

Using spectroscopic observations from the GTC, we estimate the redshift and physical properties of the \ronefourseven\ host galaxy. The FRB host is a dwarf star-forming galaxy at $z=0.130287\pm0.000002_{\mathrm{stat}}\pm0.000002_{\mathrm{sys}}$, which closely resembles the SMC in stellar mass ($\approx 4 \times 10^{8} M_{\odot}$), metallicity ($12 + \log \left( \left[ {\rm O/H} \right] \right) \approx 8.2$), SFR ($\approx0.06\,M_\odot\,\mathrm{yr^{-1}}$), and size \citep{2024A&A...681A..15M}. This characterization adds to a growing number of hyperactive FRBs, including FRB~20121102A \citep{Chatterjee_2017_Natur,Tendulkar_2017_ApJL} and FRB~20190520B \citep{Niu_2022_Natur}, that are found in dwarf galaxies similar to \ronefourseven's host. It is noteworthy that most hyperactive repeaters --- including our source, FRB~20121102A, and FRB~20190520B --- now originate from dwarfs with stellar masses $\lesssim10^9\,M_\odot$ (or luminosities $\lesssim0.01\,L^*$), a fraction significantly higher than the few percent ($\lesssim5$\,\%) of the Universe's total stellar mass found in such galaxies \citep[e.g.,][]{2012MNRAS.421..621B}. This disproportionate association can be explained if hyperactive repeaters track star-formation over stellar mass \citep{Loudas_2025_arXiv}, which can also help explain the observed prevalence of dwarf hosts. Given the small number of securely localized hyperactive FRBs, we cannot yet make a statistically robust statement on the prevalence of dwarf hosts, but the repeated association of the most active repeaters with such galaxies is suggestive and warrants future study. The association of some FRBs with dwarf galaxies may also be related to the typically low-metallicity environments found in such systems, which are conducive to the growth of high-mass stars. For example, \citet{Bassa_2017_ApJL} found the FRB~20121102A host to have a metallicity of $12 + \log \left( \left[ {\rm O/H} \right] \right) = 8.0\pm0.1$.

Our DM budget analysis for \ronefourseven\ indicates that the dominant contribution to the observed dispersion measure likely arises from the foreground halo term ($\mathrm{DM}_{\rm fhalo}$). This finding has significant implications, as this component was not explicitly modeled in previous studies \citep[e.g.,][]{Tian_2024_MNRAS, chen_ApJ_2025}, where the unusually high DM of this FRB remained only partially explained. While \cite{chen_ApJ_2025} noted the role of the foreground galaxy cluster, they reported a residual DM excess after accounting for the cluster and host H$\alpha$ emission. Our inclusion of additional intervening halos helps close this gap and suggests that a dominant foreground contribution may be a recurring feature, particularly for FRBs with unusually high extragalactic DM, as also found for FRB~20190520B \citep{2023ApJ...954L...7L}.
If this trend persists, it would support the use of foreground galaxy counts as empirical proxies for estimating the IGM contribution \citep{hsu_2025_aa}. This would not only improve our understanding of FRB environments but also enhance the precision of FRB-based cosmological measurements. Our result thus underscores the value of incorporating foreground information into FRB analyses, both for characterizing local environments and for unlocking the full potential of FRBs as cosmological probes, as previously investigated in the FLIMFLAM project \citep[e.g.,][]{2023ApJ...954L...7L,Simha_2023_ApJ}.

A major discovery in this work is the identification of the FRB host as a satellite of a more massive, star-forming spiral galaxy, DESI J212740.488+041911.64. Our GTC spectroscopic follow-up confirms that this central galaxy lies at the same redshift as the FRB host, with a projected offset of $\sim$$85$\,kpc. Until now, securely localized FRB hosts have been either isolated dwarfs or isolated spirals/ellipticals. Our system represents the first established case of an FRB residing in a gravitationally bound dwarf-central pair, adding a wholly new class of environment for FRB engines. This observed satellite-central configuration, with an SMC-like dwarf host and a more massive central galaxy, mirrors the Small Magellanic Cloud-Milky Way system, albeit at a lower overall mass scale since the central galaxy is significantly less massive than the Milky Way. This analogy is motivated by the similar stellar masses and star-formation histories of the two systems, which closely match those of the SMC-Milky Way pair. Even without clear tidal features in our current analysis, this pair invites targeted searches for interaction-driven starbursts, which are often observed in satellite-central systems. An analogous environment is seen in \citet{Gordon_2024_APJL}, where FRB~20220610A is located within a compact, interacting galaxy group at $z \approx 1$. Their DM analysis highlights a significant contribution from the group environment ($\sim 110$--$220$\,pc\,cm$^{-3}$), reinforcing the role of complex galactic systems in shaping FRB dispersion measures. Such observations could test whether tidal encounters or gravitational interactions play a significant role in creating FRB progenitors, as some authors have proposed \citep[e.g.,][]{2022ApJ...925L..20K}.

This satellite association also has profound implications for future FRB host identification and characterization. Other known repeating FRBs have shown offsets of tens of kpc from massive galaxies, such as FRB~20200120E \citep{Bhardwaj_2021_ApJL}, associated with a globular cluster within M81 \citep{Kirsten_2022_Natur}, and FRB~20240209A, offset from a quiescent elliptical galaxy \citep{Shah_2025_ApJL,Eftekhari_2025_APJL}. The case of FRB~20190208A \citep{hewitt_2024_mnras_dec}, a repeating FRB whose host appears to be an exceptionally low-luminosity ($\sim 0.001$--$0.05\,L^{*}$) dwarf galaxy offset by $\sim$$10\arcsec$ from two more massive galaxies, further illustrates this. While the lack of a redshift determination for the ultra-faint host of FRB~20190208A currently precludes a definitive association, it is likely that in the future, some FRBs will appear host-less but will be tens of arcseconds, in projection, from a massive galaxy. The latter may well prove to be the central galaxy of an as-yet unseen low-luminosity satellite host. A faint satellite like the host of \ronefourseven\ (or FRB~20190208A) would be invisible in standard follow-up at modest redshifts in archival datasets, thus masquerading as ``hostless'' or appearing offset from the nearest detected galaxy \citep[see also][]{2023MNRAS.525..994M}. The commonly used PATH algorithm \citep{Aggarwal_2021_APJ_PATH} might even assign a low probability of association to the central galaxy, given a precise and well-offset radio localization. Recognizing and systematically identifying this population of faint satellite hosts could substantially revise redshift and energetics distributions for ``hostless'' FRBs, highlighting the need for deeper optical observations around seemingly isolated FRBs.

In conclusion, our precise localization of \ronefourseven\ combined with comprehensive spectroscopic and photometric analyses reveals a complex and unique environment for a repeating FRB. We've confirmed its location within a dwarf galaxy, shed light on the potentially dominant role of foreground halo DM, and, most notably, established its first-of-its-kind satellite association within a larger galactic system. This discovery expands the known diversity of FRB host environments and underscores the importance of multi-wavelength follow-up and careful consideration of intervening structures for a complete understanding of FRB progenitors and their utility as cosmological probes. The $0.5$-kpc offset of \ronefourseven\ from its dwarf host's center is also consistent with a source that could be the product of a binary merger \citep{Nugent_2024_APJ}, although detailed space-based observations are needed to confirm the local star-formation environment. This work opens new avenues for investigating the formation channels of FRBs, particularly in interacting galactic systems, and redefines our perspective on ``hostless'' FRBs. 

\clearpage

\section{acknowledgments}\label{sec:acknowledgments}
\begin{acknowledgments}
We thank Vishwangi Shah, Alexa Gordon, and Wen-fai Fong for useful conversations. We thank Uwe Bach and Ramesh Karuppusamy for coordinating and performing observations with the 100-m Effelsberg telescope. The European VLBI Network is a joint facility of independent European, African, Asian, and North American radio astronomy institutes. Scientific results from data presented in this publication are derived from the following EVN project code(s): EK056. This work is based in part on observations with the $100$-m telescope of the MPIfR (Max-Planck-Institut f\"ur Radioastronomie) at Effelsberg (Germany) and the $32$-m radio telescope operated by the Institute of Astronomy of the Nicolaus Copernicus University in Toru\'n (Poland) and supported by a Polish Ministry of Science and Higher Education SpUB grant. The Onsala Space Observatory national research infrastructure is funded through Swedish Research Council grant No. 2017-00648. e-MERLIN is a National Facility operated by the University of Manchester at Jodrell Bank Observatory on behalf of STFC, part of UK Research and Innovation. This research has made use of the NASA/IPAC Extragalactic Database (NED), which is operated by the Jet Propulsion Laboratory, California Institute of Technology, under contract with the National Aeronautics and Space Administration. This research used data obtained with the Dark Energy Spectroscopic Instrument (DESI). DESI construction and operations is managed by the Lawrence Berkeley National Laboratory. This material is based upon work supported by the U.S. Department of Energy, Office of Science, Office of High-Energy Physics, under Contract No. DE-AC02-05CH11231, and by the National Energy Research Scientific Computing Center, a DOE Office of Science User Facility under the same contract. Additional support for DESI was provided by the U.S. National Science Foundation (NSF), Division of Astronomical Sciences under Contract No. AST-0950945 to the NSF's National Optical-Infrared Astronomy Research Laboratory; the Science and Technology Facilities Council of the United Kingdom; the Gordon and Betty Moore Foundation; the Heising-Simons Foundation; the French Alternative Energies and Atomic Energy Commission (CEA); the National Council of Humanities, Science and Technology of Mexico (CONAHCYT); the Ministry of Science and Innovation of Spain (MICINN), and by the DESI Member Institutions: www.desi.lbl.gov/collaborating-institutions. The DESI collaboration is honored to be permitted to conduct scientific research on I'oligam Du'ag (Kitt Peak), a mountain with particular significance to the Tohono O'odham Nation. Any opinions, findings, and conclusions or recommendations expressed in this material are those of the author(s) and do not necessarily reflect the views of the U.S. National Science Foundation, the U.S. Department of Energy, or any of the listed funding agencies.
M.B. is a McWilliams fellow and an International Astronomical Union Gruber fellow. M.B. also receives support from two McWilliams seed grants. M.P.S., J.W.T.H. and the AstroFlash research group acknowledge support from a Canada Excellence Research Chair in Transient Astrophysics (CERC-2022-00009); an Advanced Grant from the European Research Council (ERC) under the European Union's Horizon 2020 research and innovation programme (`EuroFlash'; Grant agreement No. 101098079); and an NWO-Vici grant (`AstroFlash'; VI.C.192.045). A.GdP. is partly supported by grant PID2022-138621NB-I00 funded by MCIN/AEI/10.13039/501100011033 and `ERDF A way of making Europe'. S.B. is supported by a Dutch Research Council (NWO) Veni Fellowship (VI.Veni.212.058). B.M. acknowledges financial support from the State Agency for Research of the Spanish Ministry of Science and Innovation, and FEDER, UE, under grant PID2022-136828NB-C41/MICIU/AEI/10.13039/501100011033, and through the Unit of Excellence Mar\'ia de Maeztu 2020--2023 award to the Institute of Cosmos Sciences (CEX2019-000918-M). A.K. acknowledges the DGAPA-PAPIIT grant IA105024. F.K. acknowledges support from Onsala Space Observatory for the provisioning of its facilities/observational support. M.P.G. acknowledges the support of the Anton Pannekoek Institute for Astronomy during his visit to the University of Amsterdam. K.N. is an MIT Kavli Fellow. N.W. is supported by the National Natural Science Foundation of China (Nos. 12288102 and 12041304)
\end{acknowledgments}

\section{Data availability}\label{sec:data_avail}

Uncalibrated visibilities of \ronefourseven and its calibration sources can be downloaded from the JIVE/EVN archive, \url{https://archive.jive.eu}, under project codes EK056A and EK056B. Filterbank files, calibrated burst visibilities, dirty maps fits files and the scripts that made Figures~\ref{fig:burst_dirty_map}, \ref{fig:burst_family_plot}, \ref{fig:elevation_radio_plot} and \ref{fig:spw_plot}, and Table~\ref{tab:burst_properties} can be accessed in our Zenodo reproduction package: {\it Zenodo DOI will be made public once the paper is accepted}. The burst baseband data (i.e. raw voltages) and the GTC spectroscopy data presented in this paper will be made available by the authors upon reasonable request.

\vspace{0mm}
\facilities{EVN, OSIRIS+(GTC)
}

\software{SFXC \citep{Keimpema_2015_ExA}, CASA \citep{CASA}, Heimdall \citep{Heimdall}, DSPSR \citep{vanStraten_2011_PASA}, FETCH \citep{Agarwal_2020_MNRAS}, AIPS \citep{AIPS}, DIFMAP \citep{difmap}, Prospector \citep{Leja2017,Prospector}.}

\appendix
\label{sec:appendix}

\section{Radio Observations}
\label{appendix:radio_observations}

\begin{figure}
    \centering
    \includegraphics[width=1.0\linewidth]{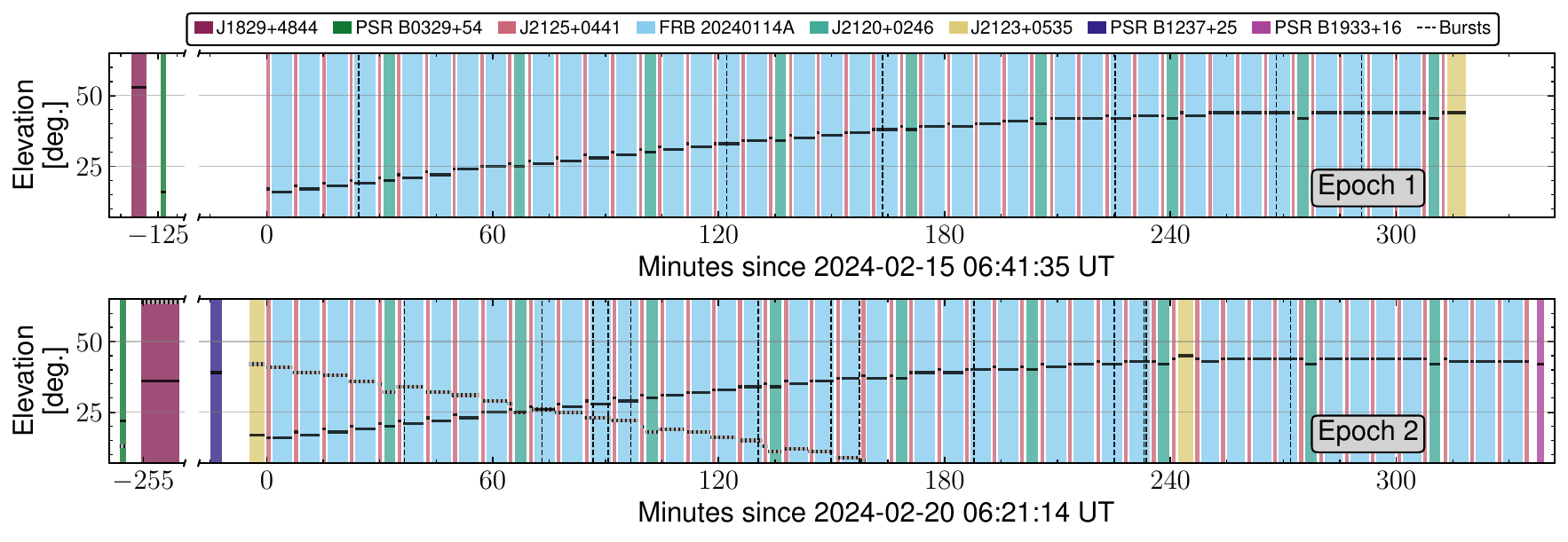}
    \caption{Observational timeline and calibration strategy for Epoch~1 (top panel; PR318A/EK056A) and Epoch~2 (bottom panel; PR319A/EK056B). The $\sim$$6$\,minute target scans on \ronefourseven are interleaved with $\sim$$2$\,minute scans on the phase-calibrator J2125+0441. Roughly twice per hour the check-source, J2120+0246, was observed for $\sim$$3.5$\,minutes per scan to verify the astrometric calibration. Other sources include J1829+4844 and J2123+0535, which are used as fringe finders and bandpass calibrators, and various pulsars to test the data quality, frequency setup and the burst-search pipeline. The solid horizontal black bars indicate the source elevation for the Effelsberg telescope (Germany) and the horizontal black-grey dashed bars indicate the same for the Tianma telescope (China). Vertical black dashed lines indicate the arrival times of the bursts. Vertical white bars indicate that the telescopes are slewing (slewing/on-source times are shown for Effelsberg, which is the slowest slewing antenna in the array).} 
    \label{fig:elevation_radio_plot}
\end{figure}

\begin{figure}
    \centering
    \includegraphics[width=0.5\linewidth]{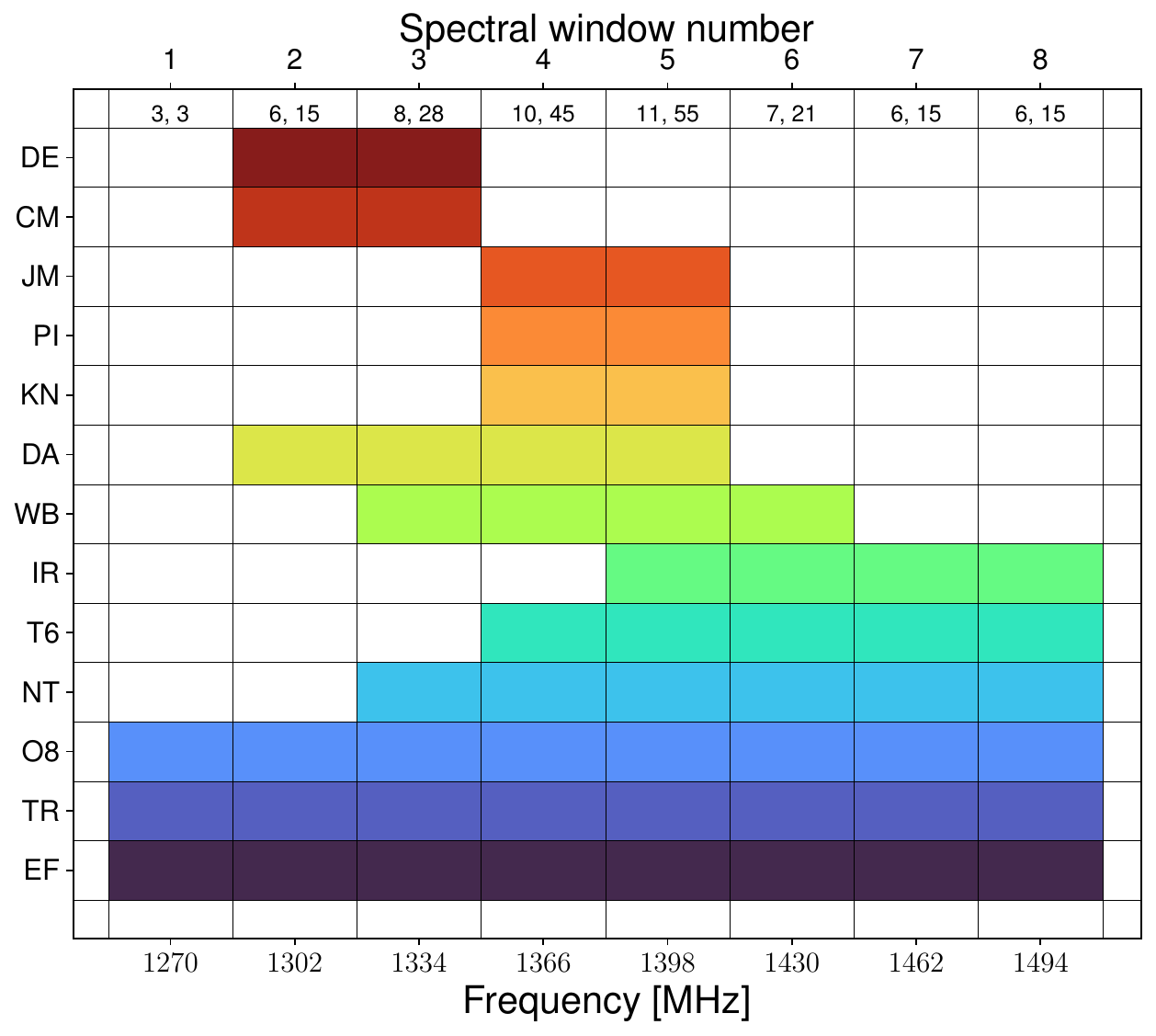}
    \caption{Observed frequencies per antenna. Every antenna recorded at least two $32$-MHz spectral windows. The numbers below the spectral window number indicate the number of antennas, $n$, that recorded that spectral window and the corresponding number of unique baselines, $\frac{1}{2}n \left( n - 1 \right) $.}
    \label{fig:spw_plot}
\end{figure}

In Figure~\ref{fig:elevation_radio_plot} we show the observational timeline and calibration strategy of our two PRECISE/EVN observations. The frequency setup of each participating telescope is illustrated in Figure~\ref{fig:spw_plot}. In Table~\ref{tab:freq_setup} we list the telescope names, abbreviations, SEFDs, recording bandwidth, and diameters.

\begin{table}
\centering
\label{tab:freq_setup}
\caption{Observational setup.}
\begin{tabular}{cccccc}
\hline
\hline
Telescope & Abbr.$\mathrm{^{a}}$ & SEFD$\mathrm{^{b}}$ & $\nu\mathrm{^{c}}$ & $\Delta\nu\mathrm{^{c}}$ & Diameter$\mathrm{^{b}}$ \\
  &  & [Jy] & [MHz] & [MHz] & [m] \\
\hline
Effelsberg                 & EF &  19 & $1254$--$1510$ & 256 & $100$   \\
Westerbork$\mathrm{^{d}}$  & WB & 420 & $1318$--$1446$ & 128 &  $25$   \\
Toru\'n                    & TR & 250 & $1254$--$1510$ & 256 &  $32$   \\
Noto$\mathrm{^{e}}$        & NT & 740 & $1318$--$1510$ & 192 &  $32$   \\
Onsala 25-m                & O8 & 310 & $1254$--$1510$ & 256 &  $25$   \\
Irbene                     & IR & 700 & $1382$--$1510$ & 256 &  $32$   \\
Tianma$\mathrm{^{f}}$      & T6 &  39 & $1350$--$1510$ & 256 &  $65$   \\
Jodrell Bank$\mathrm{^{g}}$& JM & 325 & $1350$--$1414$ & 64  & $38\times25$ \\
Cambridge$\mathrm{^{h}}$   & CM & 200 & $1286$--$1350$ & 64  & $32$    \\
Darnhall                   & DA & 450 & $1286$--$1414$ & 128 & $25$    \\
Defford                    & DE & 350 & $1286$--$1350$ & 64  & $25$    \\
Knockin$\mathrm{^{h}}$     & KN & 425 & $1350$--$1414$ & 64  & $25$    \\
Pickmere$\mathrm{^{h}}$    & PI & 400 & $1350$--$1414$ & 64  & $25$    \\
\hline
\multicolumn{6}{l}{$\mathrm{^{a}}$ Telescope abbreviation as used in Figure~\ref{fig:spw_plot}.} \\
\multicolumn{6}{l}{$\mathrm{^{b}}$ From the EVN status table.} \\
\multicolumn{6}{l}{$\mathrm{^{c}}$ The frequency range $\nu$ and the bandwidth $\Delta\nu$ do} \\
\multicolumn{6}{l}{\ \ \ not account for flagged band edges.} \\
\multicolumn{6}{l}{$\mathrm{^{d}}$ Single dish, Westerbork RT-1.} \\
\multicolumn{6}{l}{$\mathrm{^{e}}$ Recorded data up to $1574$\,MHz, but data beyond} \\
\multicolumn{6}{l}{\ \ \ $1510$\,MHz is not correlated.} \\
\multicolumn{6}{l}{$\mathrm{^{f}}$ Only participated in Epoch~2. Has a linear polarization receiver} \\
\multicolumn{6}{l}{\ \ \ and the data is converted to a circular basis before correlation.} \\
\multicolumn{6}{l}{$\mathrm{^{g}}$ Jodrell Bank Mark II.} \\
\multicolumn{6}{l}{$\mathrm{^{h}}$ Due to a technical issue these stations only recorded right} \\
\multicolumn{6}{l}{\ \ \ circular polarization (both epochs).} \\

\end{tabular}
\end{table}

\section{DM contribution of the intersecting foreground galaxies}
\label{appendix:intersection}

\begin{figure}
    \centering    
    \includegraphics[width=1.0\linewidth]{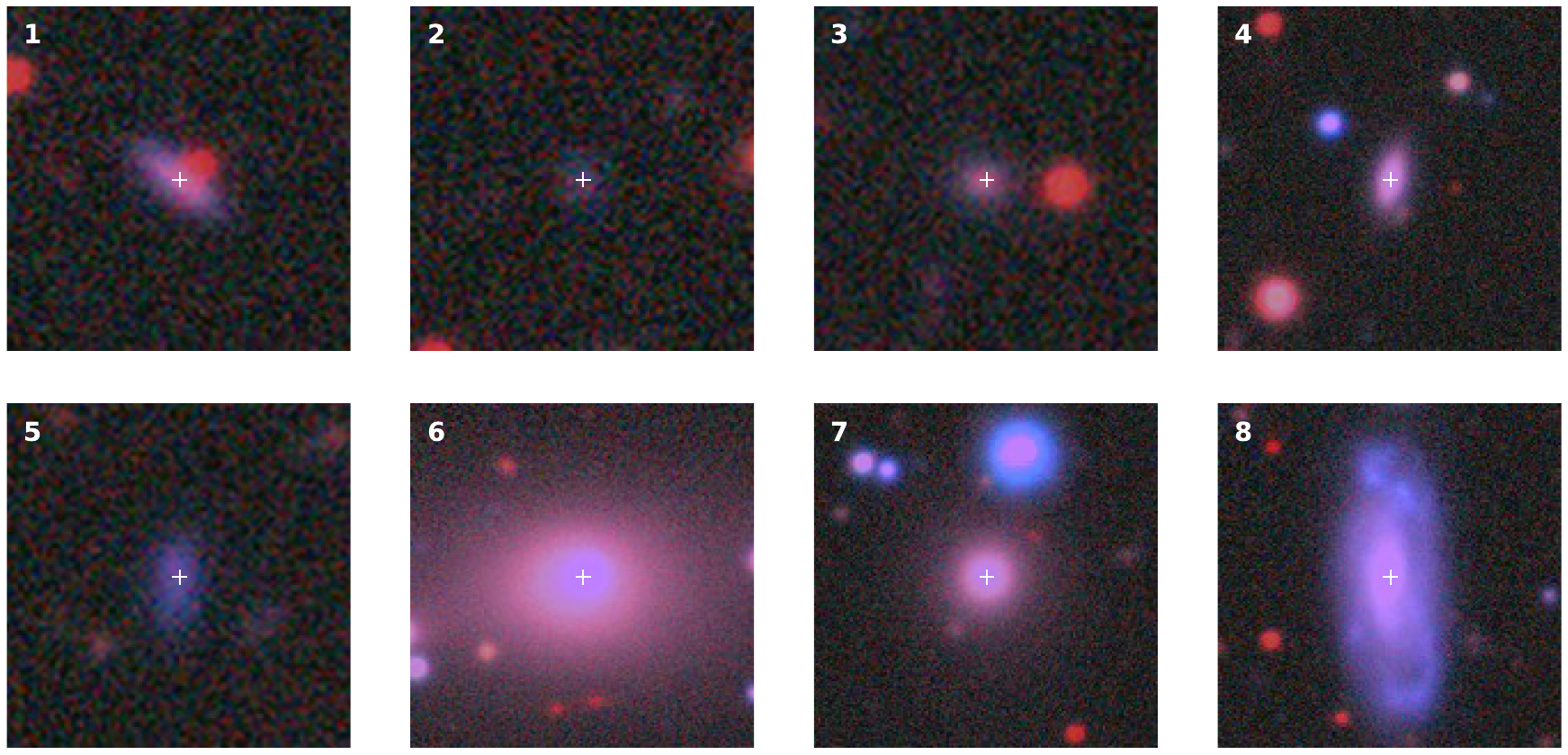}
\caption{
    RGB composite (\emph{R}=DESI $z$‑band, \emph{G}=DESI $r$‑band, \emph{B}=DESI $g$‑band) of the eight foreground galaxies whose halos intersect the \ronefourseven sightline. Each $1.5\arcmin \times 1.5\arcmin$ panel is centered on the galaxy centroid (marked by a white cross) and is displayed at the same stretch for direct comparison. Galaxies are numbered in order of increasing angular separation from the FRB position, as listed in Table~\ref{tab:frb20240114a_foreground_galaxies}.
}
\label{fig:foreground_galaxies}
\end{figure}

This section details our methodology for identifying foreground galaxies whose halos intersect the FRB sightline and contribute to its observed DM. We begin by querying the DESI Legacy Imaging Survey photometric redshift catalog \citep{2019ApJS..242....8Z} within $0.5\degr$ of the EVN-derived FRB localization. To select relevant foreground galaxies, we apply two primary criteria:
\begin{enumerate}
    \item The mean photometric redshift $\langle z \rangle$ must satisfy $\langle z \rangle - 2\,\sigma_z \le 0.130287$, where $0.130287$ is the spectroscopic redshift of the FRB host galaxy and $\sigma_z$ is the uncertainty on $\langle z \rangle$ reported in the DESI Legacy Imaging Survey photometric redshift catalog.
    \item The angular offset $\theta_{\rm offset}$ between the FRB and the galaxy centroid, normalized by the effective radius $R_{\rm eff}$, must satisfy $\theta_{\rm offset}/R_{\rm eff} > 0.018$. This ratio, hereafter referred to as the intersection ratio (or int. ratio in Table~\ref{tab:frb20240114a_foreground_galaxies}), uses $0.018$ as the empirically determined stellar-to-halo-radius ratio (SRHR) at $z\le0.1$ \citep{Somerville2018}, ensuring the FRB's projected offset falls within the foreground galaxy's virial radius ($R_{200}$). For more detail, refer \S\,\ref{sec:intervening_dm}. 
\end{enumerate}
Excluding the FRB host and DESI J212740.488+041911.64 --- i.e., the central galaxy --- these criteria yield seven candidate foreground galaxies, as shown in Figure~\ref{fig:foreground_galaxies}. Their properties are summarized in Table~\ref{tab:frb20240114a_foreground_galaxies}, and their DESI color cutouts are shown in Figure~\ref{fig:foreground_galaxies}. Note that Galaxy~1 in Figure~\ref{fig:foreground_galaxies} corresponds to the central galaxy, and its properties are discussed in \S\,\ref{section:host_properties}.

We now describe the formalism used to estimate $\mathrm{DM}_{\rm fg\,halo}$ for the interesecting galaxy. For each candidate, we first estimate the stellar mass using the empirically derived relation by \cite{2011MNRAS.418.1587T}:
\begin{equation}
\log_{10}\bigl(M_* / M_\odot\bigr)
= -0.68 + 0.70\,(g - i)\;-\;0.4 \mathrm{M}_i,
\end{equation}
where $\mathrm{M}_{i}$ is the rest-frame $i$-band AB absolute magnitude ($\mathrm{M}_{i,\odot} = 4.58$), derived from SDSS photometric data in $g$ and $i$ bands \citep{2022ApJS..259...35A}. The stellar mass is then inverted using the \cite{Moster2010} stellar-to-halo mass relation to obtain the halo mass ($M_{\rm halo}$). At the galaxy's redshift $z$, we compute the virial radius using Equation \ref{Eq:Rvir} and determine the Navarro-Frenk-White (NFW) concentration $c$ using the \cite{2014MNRAS.441.3359D} concentration-mass relation.

We then model the circumgalactic medium (CGM) gas of the galaxies using the modified NFW halo profile from \cite{2019MNRAS.485..648P}:
\begin{equation}
\rho(r)
=\frac{f_{\rm gas}\,f_b\,\rho_0}{y^{\,1-\alpha}\,(y_0 + y)^{\,2+\alpha}},
\quad
y = c\,\frac{r}{R_{200}},
\end{equation}
where $\alpha$ and $y_0$ describe the inner and outer slopes, $f_b = \Omega_b/\Omega_m \approx 0.16$ is the cosmic baryon fraction, and $f_{\rm gas}=0.75$ is the fraction of baryons in the CGM gas phase. The central density $\rho_0$ is given by:
\begin{equation}
\rho_0
=\frac{200\,\rho_c(z)}{3}\,\frac{c^3}{\ln(1+c)-c/(1+c)}.
\label{Eq:density}
\end{equation}
Following the prescription by \cite{2019MNRAS.485..648P}, we use Equation~\ref{Eq:density} to estimate the free-electron density profile $n_e(r)$ and subsequently calculate $\mathrm{DM}_{\rm halo}$ by numerically integrating $n_e(r)$ along each FRB sightline, with the measured angular offset setting the impact parameter.

Among the seven galaxies, spectroscopic redshifts for Galaxies~4, 6, 7, and 8 are available from \texttt{NED} \citep{2007ASPC..376..153M}. Using this redshift information and the formalism described above, we estimate their $\mathrm{DM}_{\rm fg\,halo}$ contributions. Notably, the projected offsets of Galaxies~4 and 8 are found to be beyond their virial radii, resulting in $\mathrm{DM}_{\rm fg\,halo} = 0$ for these two galaxies. The estimated $\mathrm{DM}_{\rm fg\,halo}$ for Galaxies~6 and 7 are reported in Table~\ref{tab:frb20240114a_foreground_galaxies}. For the remaining three galaxies (Galaxies~2, 3, and 5) that lack spectroscopic redshifts but satisfy our first selection criterion (despite $z_{\rm phot} > 0.130287$), we employ a Monte Carlo sampling approach to estimate their halo contributions. We sample the redshift $100,000$ times from a Gaussian prior $\mathcal{N} \bigl( \langle z \rangle, \sigma_{\mathrm{z}}^{2} \bigr)$. Samples with $z > 0.130287$ are excluded, effectively creating a truncated Gaussian distribution with a maximum at $z = 0.130287$. We then estimate $\mathrm{DM}_{\rm fg\,halo}$ for each redshift value. The mean of these estimated $\mathrm{DM}_{\rm fg\,halo}$ values is reported in Table~\ref{tab:frb20240114a_foreground_galaxies}. Summing all seven contributions yields a total observer-frame $\mathrm{DM}_{\rm fg\,halo}=124\,$pc\,cm$^{-3}$. If we include the DM contribution from the central galaxy's halo, estimated to be $11\,$pc\,cm$^{-3}$ based on its halo mass of $\approx 3\times10^{11}$ $M_{\odot}$ (as computed in \S\,\ref{sec:central_galaxy}) and the formalism described above, the total becomes $135\,$pc\,cm$^{-3}$. As evident from Table~\ref{tab:frb20240114a_foreground_galaxies}, Galaxy~6 contributes most significantly to the $\mathrm{DM}_{\rm fg\,halo}$ estimate; this galaxy is a bright and massive member of the galaxy cluster J212719.9+042225 \citep{2024ApJS..272...39W}. To account for systematic uncertainties inherent in the color-stellar mass calibration, halo mass inversion, concentration relation, and profile parameters, we adopt a conservative $50$\,\% one-sigma scatter on our final $\mathrm{DM}_{\rm fg\,halo}$ estimates, consistent with other similar studies \citep[see, e.g.,][]{Simha_2023_ApJ,2025arXiv250302947A}. This leads to a total observer-frame $\mathrm{DM}_{\rm fg\,halo}=135 \pm68\,$pc\,cm$^{-3}$.

Finally, to ensure no local Universe galaxies with large angular virial radii were missed by our initial $0.5\degr$ search, we cross-referenced our findings with the Heraklion Extragalactic CataloguE \citep[HECATE;][]{Kourkchi2021}, which compiles nearby galaxies within $\sim$$200$\,Mpc. Stellar masses from HECATE were converted into halo masses using the stellar-to-halo mass relation of \citet{Moster2010}, and corresponding virial radii were computed. A comprehensive search within $30\degr$ of the FRB revealed no additional foreground galaxies.

\begin{table}[ht]
\centering
\caption{Photometric and Redshift Properties of Galaxies Near \ronefourseven.} 
\label{tab:frb20240114a_foreground_galaxies}
\begin{tabular}{ccccccccccc}
\hline\hline
No. & $\alpha$ (J2000) & $\delta$ (J2000) & $r$-mag (AB) & $\langle z \rangle$ & $\sigma_z$ & $\theta_{\rm offset}$ ($''$) & Int. Ratio & $M_r$ (AB) & $z_{\rm spec}$ & DM$_{\rm fg\,halo}$ (pc\,cm$^{-3}$) \\
\midrule
2 & $21^{\mathrm{h}}27^{\mathrm{m}}38\fs57$ & $04\degr19\arcmin28\farcs9$ & 21.9 & 0.52  & 0.23  & 25.3  & 0.030 & -     & -     & 9 \\
3 & $21^{\mathrm{h}}27^{\mathrm{m}}40\fs54$ & $04\degr20\arcmin44\farcs8$ & 20.5 & 0.25  & 0.08  & 60.1  & 0.025 & -     & -     & 3 \\
4 & $21^{\mathrm{h}}27^{\mathrm{m}}40\fs87$ & $04\degr21\arcmin11\farcs5$ & 18.3 & 0.15  & 0.04  & 87.2  & 0.023 & -19.7 & 0.090 & 0 \\
5 & $21^{\mathrm{h}}27^{\mathrm{m}}42\fs82$ & $04\degr18\arcmin34\farcs5$ & 20.6 & 0.18  & 0.15  & 84.1  & 0.022 & -     & -     & 3 \\
6 & $21^{\mathrm{h}}27^{\mathrm{m}}25\fs44$ & $04\degr22\arcmin46\farcs5$ & 15.2 & 0.05  & 0.02  & 281.2 & 0.019 & -23.7 & 0.090 & 104 \\
7 & $21^{\mathrm{h}}27^{\mathrm{m}}32\fs18$ & $04\degr20\arcmin31\farcs9$ & 16.8 & 0.085 & 0.013 & 123.6 & 0.019 & -21.2 & 0.090 & 5 \\
8 & $21^{\mathrm{h}}27^{\mathrm{m}}06\fs58$ & $04\degr19\arcmin01\farcs5$ & 15.8 & 0.031 & 0.012 & 499.3 & 0.018 & -19.7 & 0.029 & 0 \\
\bottomrule
\end{tabular}
\end{table}

\section{Nebular Emission Line Analysis of the FRB Host and the Central Galaxy}
\label{sec:line_estimates}

In this section, we describe the methodology used to derive various physical properties of the FRB host and the central galaxy, namely H$\alpha$ luminosities and H$\alpha$-assisted SFR, nebular metallicity, and BPT diagnostics, using the emission lines modeled and calibrated from the GTC spectra of the FRB and the central galaxy. The procedure used to identify and calibrate the flux densities of the emission lines reported in Table~\ref{tab:flines} is described in \S\,\ref{subsec:optical_analysis}. Besides the observed ($f_{\mathrm{data}}$, $f_{\mathrm{1G}}$) and extinction-corrected ($f_{\mathrm{data,c}}$, $f_{\mathrm{1G,c}}$) fluxes and errors ($\delta$), either measured directly from the spectra or after fitting single-Gaussian and their errors, respectively, we also provide in this table the corresponding equivalent widths (EW). The extinction-corrected line ratios are given in Table~\ref{tab:ratiolines}.

\begin{table}[!ht]
    \caption{Extinction-corrected emission line ratios for the FRB host and the central galaxy.}
    \begin{tabular}{l|cccc|cccc}
        \hline
        \hline
        & \multicolumn{4}{c|}{FRB host} & \multicolumn{4}{c}{Central galaxy} \\
        Ratio & Best & $16$\,\% & $50$\,\% & $84$\,\% & Best & $16$\,\% & $50$\,\% & $84$\,\% \\
        \hline
        {[O\,\textsc{ii}]}/[O\,\textsc{iii}]   & 1.169 & 1.071 & 1.182 & 1.307 & 4.182 & 3.379 & 4.513 & 6.339 \\
        {[O\,\textsc{iii}]}/H$\beta$           & 2.826 & 2.469 & 2.806 & 3.232 & 0.710 & 0.503 & 0.695 & 0.944 \\
        {[N\,\textsc{ii}]}/H$\alpha$           & 0.077 & 0.051 & 0.078 & 0.105 & 0.262 & 0.207 & 0.260 & 0.318 \\
        {[S\,\textsc{ii}]6717}/H$\alpha$       & 0.106 & 0.077 & 0.107 & 0.139 & 0.178 & 0.136 & 0.176 & 0.222 \\
        \hline
    \end{tabular}
    \label{tab:ratiolines}
\end{table}

The line fluxes and ratios measured from our 2024 spectra are provided in Tables~\ref{tab:flines} and \ref{tab:ratiolines} and are used to compute line ratios and H$\alpha$ luminosities. For the latter, we have adopted a luminosity distance of $616$\,Mpc based on the redshift of the object ($z=0.130287$) and the cosmological parameters from \citet{Bennett_2014_ApJ}. To obtain the SFR of both targets, we made use of the calibration from \citet{Kennicutt_1998_ARAA}, for which an H$\alpha$ luminosity of $1.26 \times 10^{41}$\,erg\,s$^{-1}$ is equivalent to a SFR of $1$\,$M_{\odot}$\,yr$^{-1}$. 

The errors in the observed line fluxes were derived using the results from each individual fit to the continuum plus the corresponding line profile. The errors in the extinction-corrected line ratios, H$\alpha$ luminosities, and SFRs were computed starting from the errors in the observed line fluxes and then using a total of $1000$ Monte Carlo realizations for each measurement. The error distributions for all these measurements were assumed to be Gaussian and uncorrelated. In Table~\ref{tab:ratiolines}, we provide the output 16th--50th--84th\,\% percentiles for the line ratios, as these are not perfectly Gaussian anymore. The same quantities in the case of the H$\alpha$ luminosities and SFRs are listed in Table~\ref{tab:lha_sfr}. Thus, for the central galaxy ($A_V=1.12$; Table~\ref{tab:A-galaxy-properties}), the dust‐corrected SFR is $\mathrm{SFR}_{\rm corr} = (5.8^{+2.6}_{-0.5}\times10^{-2}\,\mathrm{M_\odot\,yr^{-1}})\times10^{0.4\times1.12} = 0.16^{+0.07}_{-0.01}\,\mathrm{M_\odot\,yr^{-1}}.$ For the FRB host ($A_V=0.11$; Table~\ref{tab:A-galaxy-properties}), we similarly obtain $\mathrm{SFR}_{\rm corr} = (5.5^{+0.4}_{-0.3}\times10^{-2}\,\mathrm{M_\odot\,yr^{-1}})\times10^{0.4\times0.11} = 0.061^{+0.004}_{-0.003}\,\mathrm{M_\odot\,yr^{-1}}.$

Next, we derive the gas-phase oxygen abundances for the FRB host and the central galaxy using the N2 and O3N2 empirical diagnostics calibrated by \citet{2013A&A...559A.114M}. The N2 diagnostic is defined as
\begin{equation}
12 + \log(\mathrm{O/H}) = 8.743 + 0.462 \times \log_{10} \left([\mathrm{N\,II}]/\mathrm{H}\alpha \right),
\end{equation}
and the O3N2 diagnostic combines [O\,\textsc{iii}]/H$\beta$ and [N\,\textsc{ii}]/H$\alpha$ ratios as:
\begin{equation}
12 + \log(\mathrm{O/H}) = 8.533 - 0.214 \times \log_{10} \left( \frac{[\mathrm{O\,III}]/\mathrm{H}\beta}{[\mathrm{N\,II}]/\mathrm{H}\alpha} \right).
\end{equation}
The systematic uncertainties in these calibrations are $0.16$\,dex and $0.18$\,dex, respectively. We combine these in quadrature with the statistical uncertainties derived from the 16th and 84th percentile values of the line-ratio posteriors to report total uncertainties.

For the FRB host galaxy, we find:
\[
12 + \log(\mathrm{O/H}) = 8.23 \pm 0.17 \quad \text{(N2)}, \qquad \log(Z/Z_\odot)_{\rm gas} = -0.46 \pm 0.17,
\]
\[
12 + \log(\mathrm{O/H}) = 8.20 \pm 0.19 \quad \text{(O3N2)}, \qquad \log(Z/Z_\odot)_{\rm gas} = -0.49 \pm 0.19.
\]
For the central galaxy:
\[
12 + \log(\mathrm{O/H}) = 8.47 \pm 0.17 \quad \text{(N2)}, \qquad \log(Z/Z_\odot)_{\rm gas} = -0.22 \pm 0.17,
\]
\[
12 + \log(\mathrm{O/H}) = 8.44 \pm 0.19 \quad \text{(O3N2)}, \qquad \log(Z/Z_\odot)_{\rm gas} = -0.25 \pm 0.19.
\]
These values assume a solar oxygen abundance of \(12 + \log(\mathrm{O/H})_\odot = 8.69\) \citep{2009ARA&A..47..481A}. Table~\ref{tab:A-galaxy-properties} lists the weighted‐mean gas‐phase metallicities for both galaxies, derived from the two independent estimates above.

\begin{figure}[ht!]
  \centering
  \includegraphics[width=\textwidth]{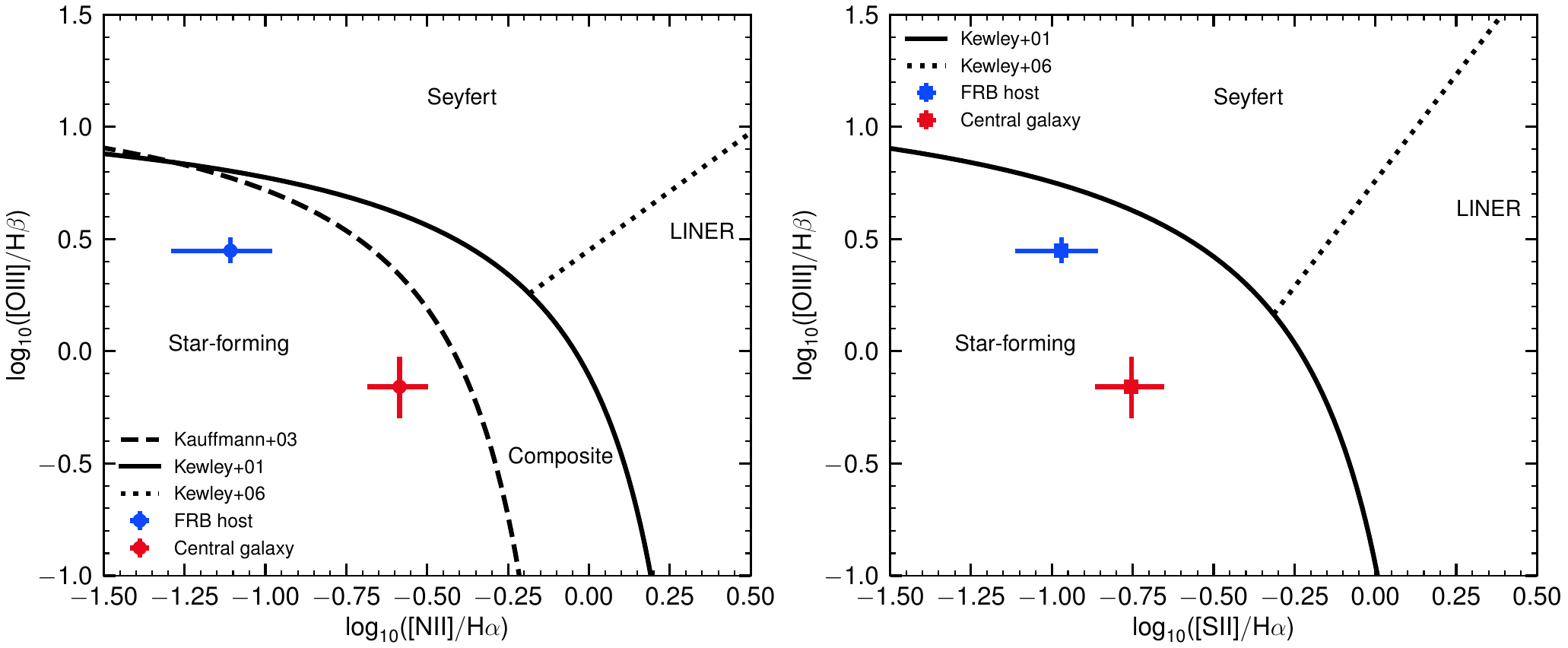}
  \caption{BPT diagnostic diagrams for the FRB host galaxy and its central galaxy.
  \textit{Left panel:} $\log_{10}([\mathrm{N\,II}]/\mathrm{H}\alpha)$ vs.\ $\log_{10}([\mathrm{O\,III}]/\mathrm{H}\beta)$, with demarcation curves from \cite[][ SF/Composite]{2003MNRAS.346.1055K}, \cite[][Composite/AGN]{2001ApJ...556..121K}, and the Seyfert-LINER division \citep{2006MNRAS.372..961K}.
  \textit{Right panel:} $\log_{10}([\mathrm{S\,II}]/\mathrm{H}\alpha)$ vs.\ $\log_{10}([\mathrm{O\,III}]/\mathrm{H}\beta)$, showing the SF/AGN curve and the Seyfert-LINER boundary from \cite{2001ApJ...556..121K} \& \cite{2006MNRAS.372..961K}, respectively.
  Error bars denote the median and $1\sigma$ uncertainties derived from the 16th, 50th and 84th percentiles of each line ratio.}
  \label{fig:BPT}
\end{figure}

To assess the dominant ionization mechanisms in both galaxies, we use the BPT diagnostic diagrams shown in Figure~\ref{fig:BPT}, employing the line ratios listed in Table~\ref{tab:ratiolines}. In the left panel, we plot $\log_{10}([\mathrm{N,II}]/\mathrm{H}\alpha)$ versus $\log_{10}([\mathrm{O,III}]/\mathrm{H}\beta)$, and in the right panel, $\log_{10}([\mathrm{S,II}]/\mathrm{H}\alpha)$ versus $\log_{10}([\mathrm{O,III}]/\mathrm{H}\beta)$. Both galaxies fall below the theoretical demarcation between AGN and star-forming galaxies from \citet[][Composite/AGN]{2001ApJ...556..121K}, the empirical star-forming sequence boundary from \citet[][SF/Composite]{2003MNRAS.346.1055K}, and the Seyfert-LINER division from \citet{2006MNRAS.372..961K}. These positions confirm that the nebular emission in both galaxies is dominated by photoionization from young stars rather than by AGN activity, thus validating the use of H,\textsc{ii}-region calibrations for metallicity diagnostics. We also compute the \([\mathrm{O\,III}]/[\mathrm{O\,II}]\) line ratio, also known as the \textit{O32} ratio, which serves as an empirical proxy for the ionization parameter in H\,\textsc{ii} regions \citep[e.g.,][]{2001ApJ...556..121K,2006A&A...459...85N}. This ratio reflects the relative abundance of doubly ionized to singly ionized oxygen and is sensitive to both the hardness of the ionizing radiation and the geometry of the gas. For the FRB host galaxy, we find \(\log_{10}([\mathrm{O\,III}]/[\mathrm{O\,II}]) = -0.07\), while the central galaxy shows a lower value of \(-0.62\). According to photoionization models \citep[e.g.,][]{2001ApJ...556..121K}, values of \(\log_{10}([\mathrm{O\,III}]/[\mathrm{O\,II}]) \gtrsim 0\) correspond to high ionization parameters (typical of extreme starbursts), whereas values between \(-1 \lesssim \log_{10}([\mathrm{O\,III}]/[\mathrm{O\,II}]) \lesssim 0\) indicate moderate ionization levels characteristic of normal star-forming galaxies. Thus, both galaxies in our system lie within the expected range for moderately ionized H\,\textsc{ii} regions, consistent with the BPT classification discussed above.

\begin{table}[!ht]
    \centering
    \caption{H$\alpha$ luminosities and SFR values for the FRB host and the central galaxy.}
    \begin{tabular}{l|cccc|cccc}
    \hline
    \hline
        & & \multicolumn{2}{c}{FRB host} & & & \multicolumn{2}{c}{Central galaxy} &  \\ \hline
        Property & Best & 16th perc. & 50th perc. & 84th perc. & Best & 16th perc. & 50th perc. & 84th perc. \\
        \hline
        H$\beta_{\text{corr}}$ [$10^{-17}$ erg\,s$^{-1}$\,cm$^{-2}$]     & 6.08 & 5.66 & 6.23 & 6.93 & 5.41 & 5.57 & 6.25 & 7.99 \\
        H$\alpha_{\text{corr}}$ [$10^{-17}$ erg\,s$^{-1}$\,cm$^{-2}$]    & 15.3 & 14.6 & 15.4 & 16.4 & 15.5 & 14.8 & 16.2 & 23.4 \\
        $L$(H$\alpha_{\text{corr}}$) [$10^{39}$ erg\,s$^{-1}$]           & 6.92 & 6.61 & 6.97 & 7.44 & 7.00 & 6.67 & 7.34 & 10.6 \\
        SFR [$10^{-2}$ $M_{\odot}$\,yr$^{-1}$]                           & 5.49 & 5.24 & 5.53 & 5.91 & 5.56 & 5.29 & 5.82 & 8.40 \\
    \hline
    \end{tabular}
    \label{tab:lha_sfr}
\end{table}

\section{Stellar Population Modeling with Prospector}
\label{sec:appendix_prospector}

We model the stellar populations of the FRB host and its central galaxy using the \texttt{prospector} SED‐fitting framework with the Flexible Stellar Population Synthesis (FSPS; \citealt{Conroy2009,Conroy2010}) library via \texttt{python‐FSPS} \citep{FSPSpython}. To simultaneously fit the broadband photometry (Table \ref{tab:sed_magnitudes}) and the GTC/OSIRIS spectroscopy, we employ dynamic nested sampling following \citet{2021ApJ...919L..24B} and \citet{Bhardwaj2024ApJb}. All data are first corrected for Milky Way extinction using the \citet{2007ApJ...663..320F} law with $A_V = 0.18$\,mag, and a $5$\,\% minimum uncertainty floor is applied to the photometry.

For the SFH we use a nonparametric ``continuity'' model with eight time bins, imposing smoothness via a Student-\(t\) prior on the log‐SFR ratios between adjacent bins \citep{Bhardwaj2024ApJb}. We assume a Kroupa initial mass function \citep{Kroupa2001} and constrain stellar metallicities with a Gaussian prior based on the mass-metallicity relation of \citet{Gallazzi2005}, using twice the observed scatter.

Dust attenuation follows the two‐component model of \citet{2000ApJ...539..718C}: young stars and nebular lines suffer a birth‐cloud optical depth \(\tau_1\), and all starlight is further reddened by a diffuse ISM optical depth \(\tau_2\) \citep{Kriek13}. We fix \(\tau_1\) as a fraction of \(\tau_2\) and impose a Gaussian prior on \(\tau_2\) whose mean and width are scaled by the line‐of‐sight path length,
$f_{\rm los} \;=\;1/\cos i$.
For the FRB host, which is likely an irregular dwarf, we assume \(i\approx0\degr\). For the central galaxy we adopt \(i=73\degr\) from the SDSS exponential‐fit axis ratio following \cite{2024Natur.634.1065B}. Thus a default face‐on prior \(\tau_2\sim\mathcal{N}(0.3,\,1.0)\) becomes $\tau_2 \;\sim\; \mathcal{N}\bigl(0.3\,f_{\rm los},\;1.0\,f_{\rm los}\bigr)$.

We account for residual sky‐line contamination with a pixel outlier model and include a two‐component mid‐infrared AGN contribution following \citet{Leja2018}. All posteriors are sampled with $500$ live points and a target sampling efficiency of $2$\,\%, and uncertainties are quoted as the 16th--84th percentile credible intervals.

\begin{table}[!ht]
\centering
\caption{Broadband photometry used for SED modeling of the central galaxy and the FRB host.}
\begin{tabular}{lccc}
\hline
\hline
Instrument & Filter & $\lambda_{\text{eff}}$ [\AA] & Magnitude [AB] \\
\hline
\multicolumn{4}{c}{Central Galaxy} \\
\hline
SDSS$^{a}$  & u   & 3543  & 21.84 $\pm$ 0.33 \\
SDSS$^{a}$  & g   & 4770  & 20.69 $\pm$ 0.06 \\
SDSS$^{a}$  & r   & 6231  & 20.03 $\pm$ 0.06 \\
SDSS$^{a}$  & i   & 7625  & 19.45 $\pm$ 0.06 \\
SDSS$^{a}$  & z   & 9134  & 19.28 $\pm$ 0.09 \\
WISE$^{b}$  & W1  & 33460 & 19.11 $\pm$ 0.06 \\
WISE$^{b}$  & W2  & 45950 & 19.48 $\pm$ 0.09 \\
\hline
\multicolumn{4}{c}{FRB Host} \\
\hline
CFIS$^{c}$  & u (MegaCam)   & 3813  & 22.95 $\pm$ 0.12 \\
SDSS$^{a}$  & g   & 4770  & 22.43 $\pm$ 0.13 \\
SDSS$^{a}$  & r   & 6231  & 21.94 $\pm$ 0.11 \\
SDSS$^{a}$  & i   & 7625  & 21.70 $\pm$ 0.16 \\
SDSS$^{a}$  & z   & 9134  & 21.64 $\pm$ 0.45 \\
\hline
\end{tabular}
\label{tab:sed_magnitudes}\\
$^{a}$ Extinction corrected model magnitude from the SDSS DR12 catalog \citep{Alam_2015_ApJS}.\\
$^{b}$From the unWISE catalog \citep{2019ApJS..240...30S}. \\
$^{c}$ From the Canada-France Imaging Survey DR3 catalog \citep[CFIS;][]{2017ApJ...848..128I} 
\end{table}

\end{document}